\documentclass[onecolumn]{IEEEtran}
\usepackage[T1]{fontenc}
\usepackage{amsmath}
\usepackage{amsthm}
\usepackage[margin=1in]{geometry}

\newtheorem{theorem}{Theorem}
\newtheorem{lemma}{Lemma}
\newtheorem{proposition}{Proposition}
\theoremstyle{definition}
\newtheorem{definition}{Definition}
\newtheorem{remark}[theorem]{Remark}

\allowdisplaybreaks[4]
\usepackage{amssymb}
\usepackage{amsfonts}
\usepackage{mathrsfs}
\usepackage{bbm}
\usepackage{graphicx} 
\usepackage{float}
\usepackage{mathtools}
\usepackage{tabularx}
\usepackage[table]{xcolor}
\usepackage{array}
\usepackage{multirow}
\usepackage{multicol}
\usepackage{cite}
\usepackage{hyperref}
\hypersetup{
	colorlinks=true,
	citecolor=blue,
	linkcolor=blue,
	urlcolor=blue
}
\usepackage{booktabs}
\definecolor{tabgray}{RGB}{242,242,242}

\title{An Explicit Below Half Threshold Bound for the Local Leakage Resilience of Shamir's Secret Sharing Scheme}

\author{Cheng Zhang, Xin Wang, and Xiangliang Kong%
\thanks{C. Zhang (czhang0707@gmail.com) and X. Wang (xinw@suda.edu.cn) are with the Department of Mathematics, Soochow University, Suzhou 215006, China.
X. Kong (xlkong@amss.ac.cn) is with the State Key Laboratory of Mathematical Sciences, Academy of Mathematics and Systems Science,
Chinese Academy of Sciences, Beijing 100190, China.}
\thanks{This work was supported in part by the National Key R\&D Program of China under Grant 2025YFA1017200 and by the National Natural Science Foundation of China under Grant No.~12601645.}
}

\begin{document}
\maketitle
	
\begin{abstract}
    As a fundamental cryptographic primitive, secret sharing distributes a secret among multiple parties and allows certain authorized parties to recover the secret while preventing unauthorized parties from knowing the secret. However, local leakage from each party member allows an adversary to learn partial information from each share, which is not guaranteed by traditional security. In this paper, we study bounds on the reconstruction threshold of Shamir's secret sharing scheme over prime fields against arbitrary 1-bit local leakage. We show that, for all sufficiently large $n$ and every prime $p>n$, the $(n,t)$-Shamir's scheme is $\exp(-\Omega(n))$-resilient whenever $t/n\geq 0.489$, for any choice of distinct nonzero evaluation points. Our proof refines the dual-code list recovery analysis and combines complementary Fourier estimates according to the support sizes and biases of the leakage functions, yielding an improved explicit reconstruction threshold bound.
\end{abstract}

\section{Introduction}
Threshold secret sharing, independently introduced by Shamir~\cite{10.1145/359168.359176} and Blakley~\cite{blakley1979safeguarding}, is a fundamental cryptographic primitive. In a threshold secret sharing scheme, any set of shares reaching the reconstruction threshold can recover the secret, whereas any smaller set reveals no information about it. However, this traditional security guarantee does not account for adversaries who obtain partial information from multiple shares. This motivates the study of leakage resilience in secret sharing schemes, particularly in Shamir's scheme, which is widely used in cryptographic applications.

In practice, side-channel attacks may reveal partial information about stored secrets or intermediate computational values, for example, through measurements of execution time or power consumption~\cite{10.5555/646761.706156,kocher1999differential}. Motivated by such attacks, Benhamouda, Degwekar, Ishai, and Rabin~\cite{benhamouda2018local} formally introduced the notion of local leakage resilience for secret sharing schemes while Goyal and Kumar~\cite{goyal2018non} independently studied a related notion. In the local leakage model, an adversary can choose any leakage function for each share and obtain a bounded number of bits from its output. Security requires that the leakage distributions corresponding to any two secrets have negligible statistical distance.

For Shamir's secret sharing scheme over prime fields, a central question is how to minimize the reconstruction threshold ratio $t/n$ while preserving security against arbitrary 1-bit local leakage. Benhamouda et al.~\cite{benhamouda2018local} established an initial bound of $0.907$, which was subsequently improved in a series of works~\cite{maji2021constructing,benhamouda2021local,maji2022improved}. Later, Klein and Komargodski~\cite{klein2023new} introduced a new Fourier-analytic approach, obtaining bounds of $0.69$ for arbitrary leakage and $0.58$ for balanced leakage functions. Kasser~\cite{Kasser2024Improvement} further improved the former bound to $0.668$. Recently, this threshold ratio bound was further improved to $0.56$ by the authors in \cite{rong-github}. 

Despite these improvements, establishing local leakage resilience against arbitrary 1-bit leakage below the reconstruction rate $t/n=1/2$ remained a major challenge. Nguyen~\cite{nguyen2024towards} established leakage resilience for almost all $1$-bit local leakage functions at any positive constant reconstruction rate, but the problem remained open for arbitrary leakage functions below half. Recently, Sun and Wootters~\cite{SunWootters2026} broke the half-barrier by exploiting connections to list decoding and list recovery of Reed--Solomon codes. Independently, Nguyen~\cite{Nguyen2026} also overcame this barrier by combining list recovery with higher-order Fourier analysis. Both works established leakage resilience for $t/n\geq 1/2-\delta$ for some constant $\delta>0$, making significant progress toward the conjecture of Benhamouda et al.~\cite{benhamouda2018local}. More recently, Hwang et al.~\cite{HMSY26} established resilience at any positive constant reconstruction rate for almost all choices of evaluation points, under additional restrictions on the field size.

Table~\ref{tab:below-half} summarizes the known bound below half-barrier for arbitrary 1-bit local leakage.

\begin{table*}[t]
\centering
\caption{Below half threshold guarantees for arbitrary 1-bit local leakage.}
\label{tab:below-half}
\footnotesize
\renewcommand{\arraystretch}{1.2}
\begin{tabularx}{\textwidth}{
@{}p{0.23\textwidth}
   p{0.15\textwidth}
   p{0.25\textwidth}
   X@{}}
\toprule
Result & Threshold rate & Evaluation points
       & Field/party regime \\
\midrule
Sun--Wootters~\cite{SunWootters2026}
& $1/2-\delta$, $\delta>0$
& Arbitrary
& Any prime $p>n$ \\

Nguyen~\cite{Nguyen2026}
& $1/2-10^{-14}$
& Arbitrary
& Any prime $p>n$ \\

Hwang et al. (quadratic)~\cite{HMSY26}
& $1/3<c<1/2$
& Nearly all valid evaluation tuples
& $n=\lfloor\lambda^u\rfloor$,
  $0<u<1/2$ \\

Hwang et al. (higher-order)~\cite{HMSY26}
& $0<c<1/2$
& Nearly all valid evaluation tuples
& $n=\lfloor\lambda^u\rfloor$,
  $1/\ell<c$,
  $0<u<\Theta_\ell^{-1}$ \\

\textbf{This work}
& $\boldsymbol{c\geq0.489}$
& \textbf{Arbitrary}
& \textbf{Any prime $p>n$} \\
\bottomrule
\end{tabularx}
\end{table*}

\subsection{Our Contribution and Main Techniques}
We consider Shamir's secret sharing scheme with $n$ parties and reconstruction threshold $t$ over a prime field $\mathbb{F}_p$. Let $\mathbf{f}=(f_1,f_2,\ldots,f_n)$ be a tuple of arbitrary 1-bit local leakage functions, where $f_i:\mathbb{F}_p\to\{-1,1\}$ is applied to the $i$-th share. For a secret $s\in\mathbb{F}_p$, the corresponding joint leakage distribution is defined as $\mathbf{f}(\mathsf{share}(s))$. Following \cite{benhamouda2018local}, Shamir's secret sharing scheme is $\varepsilon$-leakage resilient against 1-bit local leakage if, for any two secrets $s_1,s_2\in\mathbb{F}_p$, the statistical distance between the two distributions  $\mathbf{f}(\mathsf{share}(s_1))$ and $\mathbf{f}(\mathsf{share}(s_2))$ is at most $\varepsilon$. We refer to Section \ref{subsec: leakage resilient secret sharing} for the formal definitions of local leakage and leakage resilience.\vspace{0.5em}

\noindent \textbf{Main result.} We establish an improved explicit threshold bound below $1/2$ for arbitrary $1$-bit local leakage. Specifically, Shamir's secret sharing scheme is $\exp(-\Omega(n))$-resilient to arbitrary $1$-bit local leakage whenever $t/n\geq \mathbf{0.489}$.\vspace{0.5em}

Motivated by the recent results of Sun-Wootters~\cite{SunWootters2026} and Nguyen~\cite{Nguyen2026}, we develop a more refined analysis by considering both the support sizes of Fourier terms and the biases of leakage functions. By sharpening the estimates for different cases and carefully choosing the parameters, we obtain an improved explicit reconstruction threshold bound. Our result makes further progress toward the conjecture of Benhamouda et al.~\cite{benhamouda2018local}, which states that Shamir's secret sharing scheme is resilient to arbitrary $1$-bit local leakage at any positive constant reconstruction rate. On the other hand, Nielsen and Simkin~\cite{nielsen2020lower} established a $1$-bit local leakage attack at reconstruction thresholds of order $n/\log n$. Closing the gap between these bounds remains an open problem. We next outline the main techniques behind our result.\vspace{0.6em}

\noindent\textit{Refined analytic framework.} We first distinguish two cases based on the average squared bias $B$ of the leakage functions, defined as
\[B=\frac{1}{n}\sum_{i=1}^n\mu_i^2,\qquad \mu_i=\mathbb{E}_{x\leftarrow\mathbb{F}_p}[f_i(x)].\]
Let $B_0$ be a properly chosen constant. We establish two complementary bounds on the statistical distance. When $B\geq B_0$, we use an $L_2$ estimate on the differences of correlations associated with the normalized leakage functions for low supports, and an $L_1$ estimate on the correlations of the centered leakage functions for high supports. When $B<B_0$, we instead use $L_2$ estimates on the differences of correlations associated with the original leakage functions for both parts. These two estimates provide different ways to control the statistical distance, rather than treating all cases through a single estimate.\vspace{0.6em}

\noindent\textit{Low support estimates.} As in previous approaches, we use list recovery of the dual GRS code $D=C^\perp$. In contrast to the list decoding used in Nguyen's~\cite[Corollary 5]{Nguyen2026} 1-bit analysis, our argument retains the Fourier weights and applies weighted list recovery with input lists of size four. Using the MDS property together with Jeronimo's result on the list recovery of GRS codes in \cite{Jeronimo2026}, we obtain the following bound on the sum of squares of the joint correlations of the normalized leakage outputs over all subsets of $[n]$ of size $r$:
\[ \sum_{|A|=r}|\Gamma_{\mathbf{v}}(g_A)|^2 \leq \eta^{t-1}\mathsf{List}_{D}(R,4) \frac{\binom{n}{r-R}}{\binom{r}{r-R}},\]
where $\eta$ is a fixed constant, $t-1=\dim(C)$ and $r$ and $R$ are parameters to be optimized; see Theorem \ref{thm:output-decomposition} for the formal definition of $\Gamma_{\mathbf{v}}(\cdot)$. When $B\geq B_0$, this estimate directly controls the low support contributions. For $B<B_0$, we introduce a weighted tensor product operator to transfer this estimate from the normalized leakage functions to the original ones, with the additional loss controlled by $B_0$.\vspace{0.6em}

\noindent\textit{High support estimates.} When $B\geq B_0$, we apply moment estimates to the centered leakage functions, obtaining a bound that improves as $B$ increases. The main difficulty arises when $B<B_0$. Previous approaches use higher-order Fourier analysis and Gowers norms to control high support contributions. Our argument uses the structure of an MDS information set. Since only a limited number of coordinates can be strongly biased, we choose an information set containing these coordinates and divide it into two nearly equal parts. This gives a bilinear representation of the correlation. Then, applying the Cauchy--Schwarz inequality yields
\[|\Gamma_{\mathbf{v}}(f_A)|^4
\leq\mathbb{E}_{\mathbf{x},\mathbf{x'},\mathbf{y},\mathbf{y'}}\bigl[M(\mathbf{x},\mathbf{y})M(\mathbf{x'},\mathbf{y})M(\mathbf{x'},\mathbf{y'})M(\mathbf{x},\mathbf{y'})\bigr].\]
We then write the fourfold product in terms of correlations of the leakage functions. A decorrelation bound for high-rate GRS codes, together with a Gowers $U^3$ bound derived from Pollard's theorem, yield the required decay for the high support terms.\vspace{0.5em}

\noindent\textit{Combining the estimates.} We combine the low and high support estimates by choosing suitable values, including the support cutoff, bias threshold, and other parameters. For $t=\lceil0.489n\rceil$, all the resulting exponential rates are strictly negative, and this finally yields the desired threshold bound.

\subsection{Related Works}
\paragraph{Physical-bit and Hamming-weight leakage} Motivated by the probing model of Ishai, Sahai, and Wagner~\cite{ishai2003private}, several works have studied the resilience of Shamir's secret sharing scheme against physical-bit leakage, which reveals selected bits in the binary representation of each share. Maji et al.~\cite{maji2021leakage} studied this problem over prime fields, and subsequent works extended the analysis to more general finite fields~\cite{maji2024constructing} and linear code-based secret sharing schemes~\cite{nguyen2025physical}. Hamming-weight leakage, which reveals the number of ones in the binary representation of each share, has also been studied for inner-product masking over binary extension fields~\cite{BHMY26} and additive secret sharing scheme~\cite{BHMSY26}.

\paragraph{Leakage attacks and Codeword reconstruction} A complementary line of research investigates the limitations of leakage resilience through explicit attacks. Guruswami and Wootters~\cite{guruswami2016repairing,guruswami2017repairing} established connections between leakage attacks and the efficient repair of Reed--Solomon codes. Nielsen and Simkin~\cite{nielsen2020lower} showed that Shamir's scheme is vulnerable to $1$-bit local leakage when the reconstruction threshold is $O(n/\log n)$. Hoffmann and Simkin~\cite{HS23} further showed that such attacks remain effective even when a constant fraction of the leakage outputs are replaced by random noise. Other works, including those of Tamo and his collaborators, studied attacks based on Reed--Solomon codes repair and reconstruction, showing that a constant number of leaked bits per share can suffice to recover the secret in certain parameter regimes~\cite{maji2021leakage,con2022nonlinear,con2023repairing,nguyen2024towards}. More recently, Boudgoust and Simkin~\cite{cryptoeprint:2026/833} showed that leaking only $\log_2 t+O(1)$ bits from each share suffices to recover the secret in any secret sharing scheme with linear reconstruction.

\paragraph{List decoding and List recovery of Reed--Solomon codes} List decoding aims to find all codewords within a given Hamming distance of a received word, while list recovery seeks codewords that agree with prescribed lists of symbols in sufficiently many coordinates. Sudan~\cite{Sud97} introduced a list decoding algorithm for Reed--Solomon codes based on polynomial interpolation, which was later improved by Guruswami and Sudan~\cite{GS99} to reach the Johnson radius. Subsequent works made further progress beyond this radius and toward list decoding capacity~\cite{RW14,ST20,BGM23,GLSTW24,AGG25}. More recently, Brakensiek et al.~\cite{BCP26} developed an algorithm approaching list decoding capacity for arbitrary evaluation points over prime fields. Jeronimo~\cite{Jeronimo2026} established polynomial bounds on list sizes independent of the field size, including results for list recovery, while Dao et al.~\cite{DKT26} also obtained list decoding and list recovery bounds for arbitrary evaluation points.

\section{Preliminaries}
	
\textit{Notations.} Throughout this paper, $p$ is a prime number, and $[n]:=\{1,\ldots,n\}$. We use bold letters to denote vectors. For a nonempty finite set $S$, the notation $x\leftarrow S$ means that $x$ is sampled uniformly at random from $S$. We use $\mathbbm{1}_S$ to denote the indicator function of a set $S$. For a vector $\mathbf{y}$, we write $\mathsf{supp}(\mathbf{y})=\{i:y_i\neq0\}$, $\mathsf{wt}(\mathbf{y})=|\mathsf{supp}(\mathbf{y})|$. The binary entropy function is defined by $H(x):=-x\log x-(1-x)\log(1-x)$. 
	
\subsection{Secret Sharing Scheme and Generalized Reed--Solomon codes}
\begin{definition}[$(n,t)$-secret sharing]
An $(n,t)$-secret sharing scheme is defined as a randomized algorithm $\mathsf{share}$. Given a secret $s$, it generates $n$ shares, with the $i$-th party receiving $\mathsf{share}(s)_i$, such that the following hold:
\begin{itemize}
	\item[1.] \textit{Reconstruction.} Any set of at least $t$ shares can reconstruct the secret $s$.
	\item[2.] \textit{Security.} For any set of fewer than $t$ shares, their joint distribution is independent of the secret $s$.
\end{itemize}
\end{definition}

\begin{definition}[Linear codes and GRS codes]
    An $[n,k,d]_p$ linear code $C$ is a linear subspace of $\mathbb{F}_p^n$ with dimension $k$ and minimum distance $d$, i.e., $d := \min_{\mathbf{x}\in C\setminus\{\mathbf{0}\}}\mathsf{wt}(\mathbf{x})$. The dual code of $C$ is defined as $C^\perp=\{\mathbf{y}\in\mathbb{F}_p^n:\langle \mathbf{x},\mathbf{y}\rangle=0 \text{ for all }\mathbf{x}\in C\}$.
        
    Let $\alpha_1,\ldots,\alpha_n\in\mathbb{F}_p$ be distinct evaluation points, and let $v_1,\ldots,v_n\in\mathbb{F}_p^*$ be nonzero coordinate multipliers. The generalized Reed--Solomon (GRS) code of length $n$ and dimension $t$ is defined as
    $$
    \mathsf{GRS}_{\boldsymbol{\alpha},\mathbf{v}}(n,t)=\left\{(v_iQ(\alpha_i))_{i=1}^n: \deg Q<t\right\},
    $$
    where $\boldsymbol{\alpha}=(\alpha_1,\ldots,\alpha_n)$ and $\mathbf{v}=(v_1,\ldots,v_n)$. In particular, taking $v_i=1$ for all $i$ gives a Reed--Solomon code.
\end{definition}
	
\begin{definition}[Linear secret sharing scheme]
	A linear secret sharing scheme over $\mathbb{F}_p$ can be described by a generator matrix $\mathbf{G}\in\mathbb F_p^{t\times n}$ with rows $\mathbf{g}_0,\mathbf{g}_1,\ldots,\mathbf{g}_{t-1}$. To share a secret $s\in\mathbb{F}_p$, choose $\mathbf{r} =(r_1,r_2,\ldots,r_{t-1})\in\mathbb F_p^{t-1}$ uniformly at random and compute
	\[
    \mathsf{share}(s)=(s,\mathbf{r})\mathbf{G}=s\mathbf{g}_0+\sum_{i=1}^{{t-1}}r_i\mathbf{g}_i.
    \] 
    When $s=0$, the share vector is uniformly distributed over the linear code $C$ with generator matrix $\mathbf{G}$. For general $s$, $\mathsf{share}(s)$ is uniformly distributed over the affine coset $s\mathbf{g}_0+C$.
\end{definition}
	
\begin{definition}[Shamir's secret sharing scheme]
	Let $\alpha_1,\ldots,\alpha_n \in \mathbb{F}_p^{*}$ be distinct evaluation points. The $(n,t)$-Shamir’s secret sharing scheme is defined as follows:
    \begin{itemize}
        \item[1.] \textit{Sharing.} To share a secret $s\in\mathbb{F}_p$ with reconstruction threshold $t$, choose $a_1,\ldots,a_{t-1}\in \mathbb{F}_p$ independently and uniformly at random. Define the polynomial $P(X)=s+a_1X+\cdots+a_{t-1}X^{t-1}$. Then, the $i$-th party receives the $\mathsf{share}(s)_i=P(\alpha_i)$.
        \item[2.] \textit{Reconstruction.} Any $t$ shares can reconstruct the secret $s$ by Lagrange interpolation.
    \end{itemize}
        
We can write $P(X)=s+XQ(X)$ for some polynomial $Q$ of degree less than $t-1$. When $s=0$, $\mathsf{share}(0)$ is uniformly distributed over the linear code $C=\mathsf{GRS}_{\boldsymbol{\alpha},\boldsymbol{\alpha}}(n,t-1)$. For ageneral secret $s\in\mathbb F_p$, $\mathsf{share}(s)$ is uniformly distributed over the affine coset $s\mathbf{1}+C$, where $\mathbf{1}=(1,\ldots,1)\in \mathbb{F}_q^{n}$ is the all-one vector. Denote $c:=(t-1)/n$ as the rate of $C$. Since the reconstruction threshold rate is $t/n=c+1/n$, the two rates coincide asymptotically as $n\to\infty$.
\end{definition}

\subsection{Leakage Resilient Secret Sharing}\label{subsec: leakage resilient secret sharing}

\begin{definition}[Leakage distribution]
    Given a secret $s\in\mathbb{F}_p$ and a tuple of 1-bit local leakage functions $\mathbf{f}=(f_1,f_2,\ldots,f_n)$, where $f_i:\mathbb{F}_p\to\{-1,1\}$, we define the leakage distribution of $s$ by
	\[
    \mathbf{f}(\mathsf{share}(s)):=\bigl(f_1(\mathsf{share}(s)_1),f_2(\mathsf{share}(s)_2),\ldots,f_n(\mathsf{share}(s)_n)\bigr).
    \]
\end{definition}
	
Before defining leakage resilient secret sharing, we first recall the definition of statistical distance, which will be used to measure the difference between leakage distributions.
    
\begin{definition}[Statistical distance]
	For any two distributions $P$ and $Q$ over a sample space $\Omega$. Their statistical distance is defined as 
	\[
    \mathsf{SD}(P,Q):=\frac{1}{2}\sum_{x\in\Omega}\left|\Pr[P=x]-\Pr[Q=x]\right|.
    \]
\end{definition}

\begin{definition}[Local leakage resilient secret sharing scheme]
     An $(n,t)$-secret sharing scheme is $\varepsilon$-resilient against $1$-bit local leakage if, for any two secrets $s_1, s_2 \in \mathbb{F}_p$ and any tuple of $1$-bit local leakage functions $\mathbf{f}=(f_1,f_2,\ldots,f_n)$, the corresponding leakage distributions satisfy
	\[ 
    \mathsf{SD}\bigl(\mathbf{f}(\mathsf{share}(s_1)),\mathbf{f}(\mathsf{share}(s_2))\bigr) \le \varepsilon.
    \]
\end{definition}

\begin{remark}
	We also consider the setting where the adversary fully corrupts $t' < t$ parties, obtaining their shares completely, while receiving $1$-bit local leakage from the remaining parties. For Shamir's scheme, this setting can be reduced to the $(n-t',t-t')$-Shamir's scheme. For a detailed proof, we refer the reader to~\cite[Claim 4.9.1]{benhamouda2021local}.
\end{remark}

\begin{remark}[Threshold monotonicity]
    As observed in~\cite{benhamouda2018local,benhamouda2021local}, if the $(n,t)$-Shamir's scheme is $\varepsilon$-resilient against $1$-bit local leakage, then so is the $(n,t')$-Shamir's scheme for every $t\leq t'\leq n$, with the same evaluation points.
\end{remark}
	
\subsection{Fourier Analysis and Gowers Norm}
	
Throughout this paper, the Fourier analysis we use is over the prime field $\mathbb{F}_p$. For any functions $f,g:\mathbb{F}_p\to \mathbb{C}$, define $\langle f,g\rangle := \mathbb{E}_{x \leftarrow \mathbb{F}_p}[f(x)\cdot \overline{g(x)}]$, where $\overline{z}$ denotes the complex conjugate of $z\in \mathbb{C}$.
\begin{definition}[Fourier transform] 
    For $a\in\mathbb{F}_p$, define the additive character $\chi_a(x):=\exp(2\pi iax/p)$. Given a function $f:\mathbb{F}_p\to\mathbb{C}$, its Fourier coefficient at frequency $a$ is defined by $\widehat{f}(a)=\langle f,\chi_a\rangle=\mathbb{E}_{x\leftarrow \mathbb{F}_p}[f(x)\cdot \overline{\chi_a(x)}]$. The orthogonality of additive characters gives the Fourier expansion $f(x)=\sum_{a\in \mathbb{F}_p} \widehat{f}(a)\cdot \chi_a(x).$
\end{definition}
    
For a function $f:\mathbb{F}_p\to\mathbb{C}$ and $1\leq q<\infty$, define $\|f\|_q:=(\mathbb{E}_{x\leftarrow \mathbb{F}_p}\left[|f(x)|^q\right])^{1/q}$ and $\|f\|_{\infty}:=\max_{x\in\mathbb{F}_p}|f(x)|$. Then, the Parseval's identity \cite[Eq.~(4.2)]{Tao_Vu_2006} is $\|f\|_2^2=\sum_{a\in \mathbb{F}_p}|\widehat{f}(a)|^2$.
    
We next briefly recall the Gowers norms over the prime field $\mathbb{F}_p$ (see~\cite{Tao2012HigherOrder}). Since our analysis only involves real-valued functions, we omit complex conjugation in the definition.
\begin{definition}[Gowers norms]
	For a function $f:\mathbb{F}_p\to\mathbb{R}$ and an integer $d\geq 2$, the Gowers $U^d$ norm is defined by
	\[ 
    \|f\|_{U^d}^{2^d}= \mathbb{E}_{x,h_1,\ldots,h_d\leftarrow\mathbb{F}_p} \left[\prod_{\boldsymbol{\omega}\in\{0,1\}^d} f(x+\boldsymbol{\omega}\cdot\mathbf{h})\right],
    \]
	where $\mathbf{h}=(h_1,\ldots,h_d)$ and $\boldsymbol{\omega}\cdot\mathbf{h}=\sum_{j=1}^d\omega_jh_j$.
\end{definition}
In this work, we use the $U^2$ and $U^3$ norms in our estimation. The $U^2$ norm has the following Fourier representation: $\|f\|_{U^2}^4=\sum_{a\in\mathbb{F}_p}|\widehat{f}(a)|^4$ (see~\cite[p.~57, Exercise~1.3.18]{Tao2012HigherOrder}). Moreover, writing $\Delta_h f(x)=f(x)f(x+h)$, we have $\|f\|_{U^3}^8=\mathbb{E}_{h\leftarrow\mathbb{F}_p}\left[ \|\Delta_h f\|_{U^2}^4\right]$ (see~\cite[p.~57, Eq.~(1.33)]{Tao2012HigherOrder}). We distinguish between low support and high support terms according to whether $|A|\leq r_0$ or $|A|>r_0$, where $A\subseteq[n]$ and $r_0$ is a cutoff parameter specified later. The following proposition provides a uniform bound on the $U^3$ norms for any leakage functions with bounded expectation, which will be used in our analysis for leakage functions with high supports. Its proof is deferred to Appendix \ref{app:gowers}.
\begin{proposition}\label{prop:U^3 gowers}
	Let $f:\mathbb{F}_p\to\{-1,1\}$ be a leakage function satisfying $|\mathbb{E}[f]|\leq7/8$. Then
	\[\|f\|_{U^3}^8 \leq\frac{155}{384}+\frac{16}{p}+\frac{8}{3p^2}.\]
	In particular, for $p\geq 3000$, we have $\|f\|_{U^3}^8\leq\eta$, where $\eta=41/100$.
\end{proposition}

\subsection{List Recovery of GRS Codes}
Following Sun-Wotters \cite{SunWootters2026} and Nguyen \cite{Nguyen2026}, we also use List recovery bounds to control the number of dual codewords whose coordinates satisfy certain constraints.  
\begin{definition}[{\cite[List recovery list size]{Nguyen2026}}]
    For a code $C\subseteq\mathbb{F}_p^n$, an integer $0\le R\leq n$ and an input-list size $\ell\geq1$, define
    \[\mathsf{List}_C(R,\ell) :=\max_{\substack{L_1,\ldots,L_n\subseteq\mathbb{F}_p\\ |L_i|\le\ell}}\left|\left\{\mathbf{c}\in C:\left|\{i\in[n]:c_i\in L_i\}\right|\ge n-R\right\} \right|.\]
    When $\ell=1$, list recovery reduces to list decoding.
\end{definition}

The following result by Jeronimo in \cite{Jeronimo2026} on the list recovery list size of GRS codes plays a crucial role in our analysis.

\begin{theorem}[{\cite[Theorem 1.1]{Jeronimo2026}}]\label{thm:rs-list-recovery}
	Fix an integer $\ell\ge1$ and a constant $\xi\in(0,1)$. For sufficiently large $n$ and any prime $p\geq n$, let $C\subseteq\mathbb{F}_p^n$ be an Reed--Solomon code of dimension $1\le k\le(1-\xi)n$. Then
	\[ \mathsf{List}_{C}(n-k-\lceil\xi n\rceil,\ell) \le n^{O_{\ell,\xi}(1)}.\]
    The same bound holds for GRS codes by coordinatewise rescaling.
\end{theorem}
Using this list recovery result, Nguyen~\cite[Lemma 14]{Nguyen2026} derived a weighted counting bound. We use the following adapted version.
\begin{lemma}
\label{lem:symmetric-weight-list-recovery}
    Let $D\subseteq\mathbb{F}_p^n$ be a code. Fix $H\subseteq[n]$ with $|H|=j$ and an integer $j\le r\le n$. For each $i\in H$, let $w_i:\mathbb{F}_p\setminus\{0\}\to[0,\eta]$ satisfy $w_i(a)=w_i(-a),\  \sum_{a\ne0}w_i(a)\le1$, where $1/4\le\eta<1/2$. Then
    \[\sum_{\substack{\mathbf{y}\in D, \mathsf{wt}(\mathbf{y})=r\\ H\subseteq\mathsf{supp}(\mathbf{y})}}\prod_{i\in H}w_i(y_i)\leq \eta^j\mathsf{List}_D(r-j,4). \]
\end{lemma}

\subsection{Other Useful Results}\label{sec:useful-results}
We collect the properties of GRS codes and leakage functions that will be used throughout the analysis. We first introduce some properties of GRS codes and their duals.
	
\begin{lemma}[\cite{HuffmanPless2003}]\label{lem:grs-mds}
	Let $C=\mathsf{GRS}_{\boldsymbol{\alpha},\mathbf{v}}(n,t-1)$ with $2\leq t \leq n$, and set $D=C^{\perp}$. Then, any $t-1$ coordinates of a uniformly random codeword in $C$ are jointly uniform over $\mathbb{F}_p^{t-1}$. The dual code $D$ is an $[n,n-t+1,t]$ GRS code. Moreover, for every $A\subseteq[n]$ with $|A|=r\ge t$, the shortened code
	\[D_A=\{\mathbf{y}|_A:\mathbf{y}\in D, \ y_i=0\text{ for all }i\notin A\}\]
	is an $[r,r-t+1,t]$ MDS code. In particular, any $r-t+1$ coordinates of $D_A$ form an information set.
\end{lemma}
	
We use the following H\"older-type inequality to handle $t$-wise independent random variables.
\begin{lemma}\label{lem:limited-independence-holder}
	Let $Z_1,\ldots,Z_r$ be random variables such that any $t$ of them are independent, where $1\le t\le r$. Then, for arbitrary functions $f_1,\ldots,f_r$, we have
    \[
    \mathbb{E}\left[\prod_{i=1}^r|f_i(Z_i)|\right]\le\prod_{i=1}^r\|f_i(Z_i)\|_{r/t}.
    \]
\end{lemma}
	
We next bound the statistical distance between leakage distributions in terms of correlations on the corresponding affine cosets. The resulting decomposition lies at the heart of our analytical framework and allows us to separate the analysis according to whether the support sizes of the leakage functions are low or high.
	
For leakage functions $f_i:\mathbb{F}_p\to\{-1,1\}$, $i\in[n]$, write
\[
\mu_i=\mathbb E_{x\leftarrow\mathbb F_p}f_i(x),\qquad \sigma_i=\sqrt{1-\mu_i^2},\qquad B=\frac{1}{n}\sum_{i=1}^{n}\mu_i^2,\qquad h_i=f_i-\mu_i.
\]
If $f_i$ is nonconstant, define its normalized version $g_i=h_i/\sigma_i$, so that $\mathbb{E}[g_i]=0$ and $\mathbb{E}[g_i^2]=1$. If $f_i$ is constant, set $h_i=g_i=0$. For a tuple $\mathbf{f}=(f_1,\ldots,f_n)$ and $A\subseteq[n]$, put
\[
f_A(\mathbf{x})=\prod_{i\in A}f_i(x_i),\qquad f_{\emptyset}=1.
\]
	
\begin{theorem}\label{thm:output-decomposition}
	Let $C\subseteq\mathbb{F}_p^n$ be a GRS code of dimension $t-1$, and let $\mathbf{v},\mathbf{w}\in\mathbb F_p^n$. Given a collection of 1-bit local leakage functions $\mathbf f=(f_1,\ldots,f_n)$, the leakage distributions of $s_1$ and $s_2$ corresponding to the cosets $\mathbf{v}+C$ and $\mathbf{w}+C$, respectively. For $\mathbf{z}\in\{\mathbf{v},\mathbf{w}\}$ and $A\subseteq[n]$, define
	\[\Gamma_{\mathbf{z}}(f_A) =\mathbb{E}_{\mathbf{X}\leftarrow \mathbf{z}+C} \prod_{i\in A}f_i(X_i).\]
	Set $a_A^f=\Gamma_{\mathbf{v}}(f_A)-\Gamma_{\mathbf{w}}(f_A)$ and $a_A^g=\Gamma_{\mathbf{v}}(g_A)-\Gamma_{\mathbf{w}}(g_A)$. By the MDS property of $C$, we have $a_A^f=a_A^g=0$ whenever $|A|<t$. Then, for any integer $0\le r_0\le n$, the statistical distance between $\mathbf{f}(\mathsf{share}(s_1))$ and $\mathbf{f}(\mathsf{share}(s_2))$ satisfies
	\begin{align}
		\mathsf{SD}(\mathbf{f}(\mathsf{share}(s_1)),\mathbf{f}(\mathsf{share}(s_2)))\leq\frac{1}{2}\left(\sum_{\substack{A\subseteq[n]\\ t\leq|A|\leq r_0}}|a_A^g|^2\right)^{1/2}+\frac{1}{2}\sum_{\mathbf{z}\in\{\mathbf{v},\mathbf{w}\}}\sum_{\substack{A\subseteq[n]\\ |A|>r_0}} |\Gamma_{\mathbf{z}}(h_A)|. \label{eq:normalized-split}
	\end{align}
	Moreover,
	\begin{align}
		\mathsf{SD}(\mathbf{f}(\mathsf{share}(s_1)),\mathbf{f}(\mathsf{share}(s_2))) \leq \frac{1}{2} \left(\sum_{\substack{A\subseteq[n]\\t\leq|A|\leq r_0}} |a_A^f|^2\right)^{1/2} +\frac{1}{2}\left(\sum_{\substack{A\subseteq[n]\\|A|> r_0}} |a_A^f|^2\right)^{1/2}. \label{eq:walsh-split}
	\end{align}
\end{theorem}
The proof of \eqref{eq:normalized-split} uses an orthogonal expansion similar to that in Sun-Wootters~\cite[Lemma 3.3]{SunWootters2026} and Nguyen~\cite[Proposition 2]{Nguyen2026}. And \eqref{eq:walsh-split} follows from a similar expansion in the Walsh basis. For completeness, we provide their proofs in Appendix~\ref{app:statistical-distance-bounds}.
	
The two bounds involve the original leakage functions $f_i$ and their normalized versions $g_i$. The following lemma provides a uniform bound on the nonzero Fourier coefficients of both, independent of their means.
	
\begin{lemma}[{\cite[Claim 4.5]{klein2023new}}]\label{lem:leakage-fourier}
	Let $f:\mathbb{F}_p\to\{-1,1\}$ be a nonconstant leakage function. and set $g$ be its normalized version. Then, for every $a\in\mathbb F_p\setminus\{0\}$, both $\widehat{f}(a)$ and $\widehat{g}(a)$ satisfy
	\[\max\{|\widehat{f}(a)|, |\widehat{g}(a)|\} \le\frac{2}{p\sin(\pi/p)}<\gamma:=\frac{16}{25}.\]
	Moreover, $|\widehat g(a)|^2$ defines an even probability distribution on the nonzero frequencies, satisfying $|\widehat g(a)|^2\le\gamma^2=256/625<41/100=\eta$. 
\end{lemma}
	
\section{Bounds for low support}\label{sec:low-support}
	
Theorem \ref{thm:output-decomposition} bounds the statistical distance in terms of correlations indexed by supports. In this section, we use the bound on the list size for the list recovery of the dual code to estimate the contributions from normalized leakage functions. We first establish a bound for each support size and then sum over all sizes up to $r_0$.
	
\subsection{List Recovery and Low Support Estimation}\label{sec:list-recovery-low-support}
	
Let $f_i:\mathbb{F}_p\to\{-1,1\}$, $i\in[n]$, be arbitrary leakage functions, and let $g_i$ be their normalized versions as defined in Section \ref{sec:useful-results}. Then, $g_i=0$ when $f_i$ is constant.
	
The following proposition bounds the sum of squared contributions at a fixed support size. Its proof adapts the dual-code list-recovery approach of \cite{SunWootters2026,Nguyen2026}, but with a different choice of coordinate lists.
	
\begin{proposition}\label{prop:joint-low-support}
	Let $C\subseteq\mathbb{F}_p^n$ be a GRS code of dimension $t-1$, and $D=C^\perp$ be its dual code of minimum distance $t$. Then, for every $\mathbf{v}\in\mathbb{F}_p^n$ and integers $t\le r\le n$ and $1\le j\le t-1$,
	\begin{equation}
	    \sum_{\substack{A\subseteq[n]\\|A|=r}} |\Gamma_{\mathbf{v}}(g_A)|^2 \le \eta^{t-1}\mathsf{List}_D(r-j,4) \frac{\binom{n}{j}}{\binom{r}{j}}. \label{eq:joint-fixed-layer}
	\end{equation}
\end{proposition}
	
\begin{proof}
	We first assume that all leakage functions are nonconstant. Fix $H\subseteq[n]$ with $|H|=j$. For each set $A$ containing $H$ with $|A|=r$, choose $I_A$ such that $H\subseteq I_A\subseteq A$ and $|I_A|=t-1$. By the definition of $g_i$, we have $\widehat g_i(0)=\mathbb{E}[g_i]=0$. Writing $\mathbf{X}=\mathbf{v}+\mathbf{c}$, where $\mathbf{c}$ is uniformly distributed over $C$, and applying the Fourier inversion formula $g_i(x)=\sum_{a\in\mathbb{F}_p}\widehat g_i(a)\chi_a(x)$, we obtain:
	\begin{align*}
		\Gamma_{\mathbf{v}}(g_A) & =\mathbb E_{\mathbf{c}\leftarrow C}\prod_{i\in A}g_i(v_i+c_i)
		=\mathbb{E}_{\mathbf{c}\leftarrow C}\prod_{i\in A} \left(\sum_{a_i\in\mathbb F_p\setminus\{0\}} \widehat{g_i}(a_i)\chi_{a_i}(v_i+c_i)\right)\\
		& =\mathbb{E}_{\mathbf{c}\leftarrow C} \sum_{\substack{\mathbf{y}\in\mathbb{F}_p^n\\ \mathsf{supp}(\mathbf{y})=A}} \left(\prod_{i\in A}\widehat{g_i}(y_i)\right) \chi_1(\langle\mathbf{v},\mathbf{y}\rangle)\chi_1(\langle\mathbf{c},\mathbf{y}\rangle)\\
		& =\sum_{\substack{\mathbf{y}\in\mathbb F_p^n\\ \mathsf{supp}(\mathbf{y})=A}} \chi_1(\langle\mathbf{v},\mathbf{y}\rangle) \left(\prod_{i\in A}\widehat{g_i}(y_i)\right) \mathbb{E}_{\mathbf{c}\leftarrow C}\chi_1(\langle\mathbf{c},\mathbf{y}\rangle)\\
		& =\sum_{\substack{\mathbf{y}\in D\\ \mathsf{supp}(\mathbf{y})=A}} \chi_1(\langle\mathbf{v},\mathbf{y}\rangle) \left(\prod_{i\in A}\widehat{g_i}(y_i)\right).
	\end{align*}
	By Lemma \ref{lem:grs-mds}, the set $A\setminus I_A$ is an information set for the shortening of $D$ to $A$. Thus, every codeword $\mathbf{y}\in D$ with $\mathsf{supp}(\mathbf{y})\subseteq A$ is uniquely determined by its coordinates on $A\setminus I_A$. Applying Cauchy--Schwarz to the preceding Fourier expansion, we obtain
	\begin{align*}
		|\Gamma_{\mathbf{v}}(g_A)|^2
		& \leq\left(\sum_{\substack{\mathbf{y}\in D\\ \mathsf{supp}(\mathbf{y})=A}} \left|\chi_1(\langle\mathbf{v},\mathbf{y}\rangle) \prod_{i\in I_A}\widehat{g_i}(y_i)\right|^2\right) \cdot \left(\sum_{\substack{\mathbf{y}\in D\\ \mathsf{supp}(\mathbf{y})=A}} \left|\prod_{i\in A\setminus I_A} \widehat{g_i}(y_i)\right|^2\right)\\
		& = \left(\sum_{\substack{\mathbf{y}\in D\\\mathsf{supp}(\mathbf{y})=A}} \prod_{i\in I_A}\left|\widehat{g_i}(y_i)\right|^2\right) \left(\sum_{\substack{\mathbf{y}\in D\\\mathsf{supp}(\mathbf{y})=A}} \prod_{i\in A\setminus I_A}\left|\widehat{g_i}(y_i)\right|^2\right)\\
		& \leq\left(\sum_{\substack{\mathbf{y}\in D\\\mathsf{supp}(\mathbf{y})=A}} \prod_{i\in I_A}\left|\widehat{g_i}(y_i)\right|^2\right)\left(\sum_{(y_i)_{i\in A\setminus I_A}\in (\mathbb{F}_p^*)^{r-t+1}} \prod_{i\in A\setminus I_A}\left|\widehat{g_i}(y_i)\right|^2\right)\\
		& = \sum_{\substack{\mathbf{y}\in D\\\mathsf{supp}(\mathbf{y})=A}}
			\prod_{i\in I_A}\left|\widehat{g_i}(y_i)\right|^2\cdot\prod_{i\in A\setminus I_A}\left(\sum_{y_i\neq0}\left|\widehat {g_i}(y_i)\right|^2\right)\\
			&\leq \eta^{t-1-j} \sum_{\substack{\mathbf{y}\in D\\\mathsf{supp}(\mathbf{y})=A}} \prod_{i\in H}\left|\widehat{g_i}(y_i)\right|^2.
	\end{align*}
    Here, the last inequality follows from $\sum_{y_i\ne0}|\widehat{g_i}(y_i)|^2=1$, $\left|\widehat{g_i}(y_i)\right|^2\le\eta$ together with $|I_A\setminus H|=t-1-j$. Summing over all $r$-subsets $A$ containing the fixed set $H$, we obtain
	\begin{equation}
		\sum_{\substack{A\supseteq H\\|A|=r}} |\Gamma_{\mathbf{v}}(g_A)|^2 \le \eta^{t-1-j} \sum_{\substack{\mathbf{y}\in D,\ \mathsf{wt}(\mathbf{y})=r\\ H\subseteq\mathsf{supp}(\mathbf{y})}} \prod_{i\in H}|\widehat g_i(y_i)|^2. \label{eq:joint-support-fourier-bound}
	\end{equation}
	By the preceding Fourier bound and Parseval's identity, we have $|\widehat{g_i}(-a)|^2 = |\widehat{g_i}(a)|^2,~\sum_{a\ne 0}|\widehat{g_i}(a)|^2 = 1$, and $0\le |\widehat{g_i}(a)|^2\le \eta$ by Lemma \ref{lem:leakage-fourier}. Thus, the weights $w_i(a):=|\widehat g_i(a)|^2$, $a\in\mathbb{F}_p\setminus\{0\}$, satisfy the conditions of Lemma~\ref{lem:symmetric-weight-list-recovery}. Applying this lemma to \eqref{eq:joint-support-fourier-bound}, for every fixed $H$, we have
	\[ \sum_{\substack{A\supseteq H\\|A|=r}} |\Gamma_{\mathbf{v}}(g_A)|^2 \le \eta^{t-1-j}\eta^j \mathsf{List}_D(r-j,4)= \eta^{t-1}\mathsf{List}_D(r-j,4).\]
	Finally, summing over all $j$-subsets $H\subseteq[n]$ and interchanging the order of summation, we obtain
	\begin{align*}
		\binom{n}{j}\eta^{t-1}\mathsf{List}_D(r-j,4) \geq\sum_{\substack{H\subseteq[n]\\|H|=j}} \sum_{\substack{A\supseteq H\\|A|=r}} |\Gamma_{\mathbf{v}}(g_A)|^2 =\sum_{\substack{A\subseteq[n]\\|A|=r}} \sum_{\substack{H\subseteq A\\|H|=j}} |\Gamma_{\mathbf{v}}(g_A)|^2 =\binom{r}{j}\sum_{|A|=r}|\Gamma_{\mathbf{v}}(g_A)|^2.\\
	\end{align*}
Dividing both sides by $\binom{r}{j}$ gives the desired bound.
		
If some leakage function $f_i$ is constant, then $g_i\equiv0$. Thus, $\Gamma_{\mathbf{v}}(g_A)=0$ whenever $A$ contains such a coordinate. For all remaining subsets $A$, the preceding argument applies unchanged. Hence, the bound also holds in the presence of constant leakage functions. 
\end{proof}
	
Proposition \ref{prop:joint-low-support} bounds the squared normalized correlations at each support size. We now sum these bounds over all sizes up to $r_0$. This result will also be used to estimate the low support terms for general leakage functions.
	
\begin{lemma}\label{lem:normalized-low-support}
	Under the assumptions of Proposition \ref{prop:joint-low-support}, fix a constant $\zeta\in(0,1)$ and set $ R=t-1-\lceil\zeta n\rceil, r_0=t-1+R$. Assume $1\le R \le n$ and $r_0\le n$. Then,
	\begin{equation}
		\sum_{\substack{A\subseteq[n]\\|A|\le r_0}}|a_A^g|^2 \leq n^{O(1)}\exp\bigl(nE_{\mathsf{low}}(c,a)\bigr), \label{eq:total-low-support-exponent}
	\end{equation}
	where $c=(t-1)/n$, $a=R/n$ and
	\[E_{\mathsf{low}}(c,a)=c\log\eta+(1+a)\log2-(1+a)\log(1+a)+a\log a.\]
\end{lemma}
	
\begin{proof}
	The dual code $D$ is a GRS code of dimension $k=n-t+1$. Since $R=t-1-\lceil\zeta n\rceil\ge1$, we have $k\leq(1-\zeta)n$ and $n-k-\lceil\zeta n\rceil=R$. Thus, applying Theorem \ref{thm:rs-list-recovery} with $\ell=4$ and $\xi=\zeta$, yields $ \mathsf{List}_D(R,4)\le n^{O(1)}$. For $t\leq r\le r_0$, set $j=r-R$, by Proposition \ref{prop:joint-low-support} and the inequality $|x-y|^2\le 2|x|^2+2|y|^2$, we obtain
	\begin{equation}
		\sum_{|A|=r}|a_A^g|^2 =\sum_{|A|=r}\left|\Gamma_{\mathbf{v}}(g_A)-\Gamma_{\mathbf{w}}(g_A)\right|^2 \le4\eta^{t-1}\mathsf{List}_D(R,4)\frac{\binom{n}{r-R}}{\binom{r}{r-R}}.
		\label{eq:low-support-layer}
	\end{equation}
	The binomial ratio attains its maximum when $r$ is an integer nearest to $(n+R)/2$. Setting $a=R/n$, Stirling's formula gives 
    \[\frac{\binom{n}{r-R}}{\binom{r}{r-R}} \leq n^{O(1)}\exp\left(n\left((1+a)\log2-(1+a)\log(1+a)+a\log a\right)\right).\]
	Combining this estimate with \eqref{eq:low-support-layer}, and summing over $t\le r\le r_0$, we obtain
	\[ \sum_{\substack{A\subseteq[n]\\|A|\le r_0}}|a_A^g|^2 \le n^{O(1)} \exp\bigl(nE_{\mathsf{low}}(c,a)\bigr). \]
	This completes the proof.
\end{proof}

\subsection{Reduction to General Leakage}\label{sec:general-leakage}

Lemma \ref{lem:normalized-low-support} bounds the coefficients $a_A^g$ associated with normalized leakage. To use \eqref{eq:walsh-split} in Theorem \ref{thm:output-decomposition}, we need an estimate for $a_A^f$ instead. The following lemma establishes this bound, with the loss controlled by the average squared bias.
	
\begin{lemma}\label{lem:normalized-low-support-transfer}
	Consider a GRS code $C\subseteq\mathbb F_p^n$ of dimension $t-1$ and $f_1,\ldots,f_n:\mathbb{F}_p\to\{-1,1\}$ be leakage functions with normalized versions $g_i$ as in Section~\ref{sec:useful-results}. Suppose that $ B\leq B_0\le1$. Then, for every $1<u<\sqrt2$ and every integer $0\le r_0\le n$,
	\begin{equation}
        \sum_{\substack{A\subseteq[n]\\|A|\le r_0}} |a_A^f|^2 
        \le u^{-2t}\lambda_{u^2}(B_0)^{n} \sum_{\substack{A\subseteq[n]\\|A|\le r_0}} |a_A^g|^2,
		\label{eq:normalised-low-support-transfer}
	\end{equation}
	where, for $1<T<2$ and $0\le x\le1$,
	\[ \lambda_T(x) =\frac{1+T+(1-T)x+\sqrt{(T-1)^2(1-x)^2+4x}}{2}.\]
\end{lemma}

\begin{proof}
	Recall that $f_i=\mu_i+\sigma_i g_i$, where $\sigma_i=\sqrt{1-\mu_i^2}$. Expanding the product, for each $\mathbf{z}\in\{\mathbf{v},\mathbf{w}\}$, we obtain
	\[\Gamma_{\mathbf{z}}(f_A)=\mathbb{E}_{\mathbf{X}\leftarrow \mathbf{z}+C}\prod_{i\in A}(\mu_i+\sigma_i g_i(X_i))=\sum_{S\subseteq A} \left(\prod_{i\in A\setminus S}\mu_i\right) \left(\prod_{i\in S}\sigma_i\right) \Gamma_{\mathbf{z}}(g_S).\]
	Then, by the definition of $a_A^f$ in Theorem \ref{thm:output-decomposition}, this leads to
	\begin{equation}
		a_A^f=\Gamma_{\mathbf{z}}(f_A)-\Gamma_{\mathbf{w}}(f_A) =\sum_{S\subseteq A} \left(\prod_{i\in A\setminus S}\mu_i\right) \left(\prod_{i\in S}\sigma_i\right)a_S^g.
		\label{eq:coefficient-change}
	\end{equation}
	This identity also holds for constant $f_i$, for which $\sigma_i=g_i=0$ by convention.
		
	Let $\mathbf{a}^f=(a_A^f)_{A\subseteq[n]}$ and $\mathbf{a}^g=(a_A^g)_{A\subseteq[n]}$ be vectors in $\mathbb{R}^{2^n}$, indexed by the indicator vectors of subsets $A\subseteq[n]$. Then \eqref{eq:coefficient-change} can be written as
    $$
    \mathbf{a}^f=\mathbf{M}\mathbf{a}^g,
    $$
    where
	$$
    \mathbf{M}=\bigotimes_{i=1}^n \mathbf{M}_i,\qquad
	\mathbf{M}_i=\begin{pmatrix}
		1&0\\ \mu_i&\sigma_i
		\end{pmatrix}.
    $$
	For a vector $\mathbf{a}=(a_A)_{A\subseteq[n]}$, define the coordinate projection $P_{\le r_0}$ by
	\[
	(P_{\le r_0}\mathbf{a})_A=
	\begin{cases}
		a_A, & |A|\le r_0,\\
		0, & |A|>r_0.
	\end{cases}
    \]
	Since $\mathbf{M}_{A,S}=0$ unless $S\subseteq A$, we have $P_{\le r_0}\mathbf{M}=P_{\le r_0}\mathbf{M}P_{\le r_0}$. Set $\mathbf{z}=P_{\le r_0}\mathbf{a}^g$. Then $z_A=0$ unless $t\leq |A|\leq r_0$.
		
	For $1<u<\sqrt{2}$, define the diagonal degree-weighting operator $\mathbf{D}_u$ by
	\[ (\mathbf{D}_u\mathbf{a})_A=u^{|A|}a_A,\qquad \mathbf{D}_u=\bigotimes_{i=1}^n\text{diag}(1,u).\]
	Since $z_A=0$ unless $t\le |A|\le r_0$, we have
	\[ \|\mathbf{D}_u^{-1}\mathbf{z}\|_2^2 =\sum_{t\le|A|\le r_0}u^{-2|A|}|z_A|^2 \le u^{-2t}\|\mathbf{z}\|_2^2.\]
	Let $\|\cdot\|_{\text{op}}$ denote the induced operator norm on matrices. Then, by the multiplicativity of operator norms under Kronecker products, we have
	\[ \mathbf{M}\mathbf{D}_u=\bigotimes_{i=1}^n\bigl(\mathbf{M}_i\text{diag}(1,u)\bigr), \qquad \|\mathbf{M}\mathbf{D}_u\|_{\text{op}} =\prod_{i=1}^n\|\mathbf{M}_i\text{diag}(1,u)\|_{\text{op}}. 
    \]
	Using $P_{\le r_0}\mathbf{M}=P_{\le r_0}\mathbf{M}P_{\le r_0}$, and $\|P_{\le r_0}\|_{\text{op}}\le1$ by the definition of operator norm, we obtain
	\begin{align}
		\|P_{\le r_0}\mathbf{a}^f\|_2
		& =\|P_{\le r_0}\mathbf{M}\mathbf{a}^g\|_2
		   =\|P_{\le r_0}\mathbf{M}P_{\le r_0}\mathbf{a}^g\|_2
		   =\|P_{\le r_0}\mathbf{M}\mathbf{z}\|_2\notag\\
		& =\|P_{\le r_0}\mathbf{M}\mathbf{D}_u\mathbf{D}_u^{-1}\mathbf{z}\|_2 \le\|\mathbf{M}\mathbf{D}_u\|_{\text{op}}\|\mathbf{D}_u^{-1}\mathbf{z}\|_2\notag\\
		& \le u^{-t} \left(\prod_{i=1}^n \|\mathbf{M}_i\text{diag}(1,u)\|_{\mathrm{op}}\right) \|P_{\le r_0}\mathbf{a}^g\|_2.
		\label{eq:weighted-low-support-transfer}
	\end{align}
	By Lemma \ref{lem:operator-norm} in Appendix \ref{app:normalized-low-support-transfer}, for $T=u^2\in(1,2)$, we have
	\[\prod_{i=1}^n\|\mathbf{M}_i\text{diag}(1,u)\|_{\text{op}}  \le\lambda_{u^2}(B)^{n/2} \le\lambda_{u^2}(B_0)^{n/2}.\]
	Substituting this bound into~\eqref{eq:weighted-low-support-transfer} proves~\eqref{eq:normalised-low-support-transfer}.
\end{proof}
	
\section{Bounds for High Supports}\label{sec:high-support}
In Section \ref{sec:low-support}, we establish bounds for supports of size at most $r_0$. We now consider the high support terms with $|A|>r_0$. We first bound the correlations of arbitrary leakage functions on fixed supports, and then use the average squared bias to control their total contribution over high supports.
	
\subsection{High Support Control with Small Bias}\label{sec:high-support-small-bias}
We first establish a correlation bound for Boolean functions over high-rate GRS codes. The proof is given in  Appendix~\ref{lem:decor proof}.
\begin{lemma}\label{lem:decor}
	Let $C\subseteq\mathbb{F}_p^m$ be a dimension-$d$ GRS code with $d\le m\le 11d/10$ and $p\ge\max(m,3000)$. For arbitrary Boolean functions $f_i:\mathbb{F}_p\to\{-1,1\}$,
	\[
	\left|\mathbb{E}_{\mathbf{X}\leftarrow C}\prod_if_i(X_i)-\prod_i\mathbb{E}f_i\right|
	\le m^{O(1)}\gamma^d,\qquad \gamma=16/25.
	\]
	The same holds for affine translates. If $d=m$, the error is zero.
\end{lemma}

We use this bound to control the high support terms when the average squared bias is small.
	
\begin{proposition}\label{prop:large-support-small-bias}
	Let $p\ge3000$ be a prime and $3\le t\le n$. Consider a GRS code $C\subseteq\mathbb F_p^n$ of dimension $t-1$ and leakage functions $f_1,\ldots,f_n:\mathbb F_p\to\{-1,1\}$. Choose $A\subseteq[n]$ with $|A|=r$ and $J\subseteq A$ with $|J|=t-1$, then put $I=A\setminus J$ with $|I|=m$. Suppose that
	\[ \left\lfloor\frac{t-1}{2}\right\rfloor\le m
	\le\frac{11}{10}(t-1), \qquad |\mathbb{E} f_i|\le\frac{7}{8}\quad(i\in I).\]
	Then, for every $\mathbf{v}\in \mathbb{F}_p^n$, we have
	\begin{equation}
		|\Gamma_{\mathbf{v}}(f_A)| \le\eta^{\frac{1}{8}\min\{m,\,2\lfloor(t-1)/2\rfloor\}} +r^{O(1)}\gamma^{(t-1)/4},
		\notag
		\label{eq:large-support-small-bias}
	\end{equation}
	where $\gamma=16/25$ and $\eta=41/100$.
\end{proposition}

\begin{proof}
	For each $i\in[n]$, define $\widetilde{f_i}(x)=f_i(x+v_i)$. Then $\Gamma_{\mathbf{v}}(f_A)=\Gamma_{\mathbf{0}}(\widetilde f_A)$. Since translation preserves Booleanity, the $U^3$ norm, and all bounds used below, it suffices to prove the bound for $\mathbf{v}=\mathbf{0}$. We retain the notation $f_i$ for the translated functions.
		
	Put $a=\lfloor(t-1)/2\rfloor$ and $b=t-1-a$, and partition $J=J_1\sqcup J_2$ with $|J_1|=a$ and $|J_2|=b$. Since $J$ is an information set, a uniform codeword is determined by independent uniform vectors $\mathbf{x}\in\mathbb{F}_p^{J_1}$ and $\mathbf{y}\in\mathbb{F}_p^{J_2}$. For each $i\in I$, let $\ell_i(\mathbf{x},\mathbf{y})$ denote the $i$-th coordinate of the codeword, and define
	\[G_1(\mathbf{x})=\prod_{i\in J_1}f_i(x_i),\qquad G_2(\mathbf{y})=\prod_{i\in J_2}f_i(y_i),\qquad M(\mathbf{x},\mathbf{y})=\prod_{i\in I}f_i\bigl(\ell_i(\mathbf{x},\mathbf{y})\bigr).\]
    Since $J$ is an information set, the definitions of $G_1$, $G_2$, and $M$ yields that
    \[ 
    \Gamma_{\mathbf{0}}(f_A)= \mathbb E_{\mathbf{x},\mathbf{y}} G_1(\mathbf{x})M(\mathbf{x},\mathbf{y})G_2(\mathbf{y}),
    \]
    where $\mathbf x\in\mathbb F_p^{J_1}$ and $\mathbf y\in\mathbb F_p^{J_2}$ are independent and uniform.
    
    Let $\mathbf{x}'$ and $\mathbf{y}'$ be independent copies of $\mathbf{x}$ and $\mathbf{y}$, respectively. We first apply the Cauchy--Schwarz inequality. Since $G_2(\mathbf y)^2=1$, we have
    \begin{align*}
    |\Gamma_{\mathbf 0}(f_A)|^2 &=
    \left| \mathbb E_{\mathbf y} G_2(\mathbf y) \mathbb E_{\mathbf x} G_1(\mathbf x)M(\mathbf x,\mathbf y)
    \right|^2 \\
    & \le \mathbb E_{\mathbf y}
    \left( \mathbb E_{\mathbf x} G_1(\mathbf x)M(\mathbf x,\mathbf y) \right)^2 = \mathbb E_{\mathbf x,\mathbf x'} G_1(\mathbf x)G_1(\mathbf x') \mathbb E_{\mathbf y} M(\mathbf x,\mathbf y)M(\mathbf x',\mathbf y).
    \end{align*}
    Applying Cauchy--Schwarz inequality again and using $G_1(\mathbf{x})^2=1$, we obtain
    \begin{equation}\label{eq:prop3-four-point-cs}
    |\Gamma_{\mathbf 0}(f_A)|^4 \le \mathbb E_{\mathbf x,\mathbf x'} \left( \mathbb E_{\mathbf y} M(\mathbf x,\mathbf y)M(\mathbf x',\mathbf y) \right)^2 = 
    \mathbb E_{\mathbf x,\mathbf x',\mathbf y,\mathbf y'}
    \left[M(\mathbf x,\mathbf y)M(\mathbf x',\mathbf y)M(\mathbf x',\mathbf y')M(\mathbf x,\mathbf y')\right]=:
    \mathcal Q.
    \end{equation}
    In particular, $\mathcal Q\ge 0$.
    
    We now estimate $\mathcal Q$. Write $\Delta\mathbf{x}=\mathbf{x'}-\mathbf {x}$, $\Delta\mathbf{y}=\mathbf{y'}-\mathbf{y}$. Clearly, there is a one-to-one correspondence between $(\mathbf x,\mathbf x',\mathbf y,\mathbf y')$ and $(\mathbf x,\Delta\mathbf x,\mathbf y,\Delta\mathbf y)$. Consequently, $\mathbf x,\mathbf y, \Delta\mathbf x,\Delta\mathbf y$ are independent uniform vectors in their respective spaces. For each $i\in I$, put $\delta_i=\ell_i(\Delta\mathbf x,\mathbf 0)$, $\theta_i=\ell_i(\mathbf 0,\Delta\mathbf y)$, and write $\boldsymbol\delta=(\delta_i)_{i\in I}$ and $\boldsymbol\theta=(\theta_i)_{i\in I}$. For fixed $\boldsymbol\delta,\boldsymbol\theta$, define
    \[
    q_{i,\boldsymbol\delta,\boldsymbol\theta}(z) = f_i(z)f_i(z+\delta_i) f_i(z+\theta_i)f_i(z+\delta_i+\theta_i).
    \]
    By the linearity of $\ell_i$, we have $\ell_i(\mathbf{x}',\mathbf{y})=\ell_i(\mathbf{x},\mathbf{y})+\delta_i$, $\ell_i(\mathbf{x},\mathbf{y}')=\ell_i(\mathbf{x},\mathbf{y})+\theta_i$, and $\ell_i(\mathbf{x}',\mathbf{y}')=\ell_i(\mathbf{x},\mathbf{y})+\delta_i+\theta_i$. Consequently, the product of the four factors of $M$ in $\mathcal Q$ can be written as
    \begin{align*}
    & M(\mathbf{x},\mathbf{y})M(\mathbf{x}',\mathbf{y})M(\mathbf{x}',\mathbf{y}')M(\mathbf{x},\mathbf{y}') \\
     = & \prod_{i\in I}
    \Bigl[
    f_i\bigl(\ell_i(\mathbf{x},\mathbf{y})\bigr)f_i\bigl(\ell_i(\mathbf{x},\mathbf{y})+\delta_i\bigr) f_i\bigl(\ell_i(\mathbf{x},\mathbf{y})+\theta_i\bigr)f_i\bigl(\ell_i(\mathbf{x},\mathbf{y})+\delta_i+\theta_i\bigr)
    \Bigr]\\
     = & \prod_{i\in I} q_{i,\boldsymbol{\delta},\boldsymbol{\theta}} \bigl(\ell_i(\mathbf{x},\mathbf{y})\bigr).
    \end{align*}
    
    Recall that $(\mathbf{x},\mathbf{y})$ is independent of $(\Delta\mathbf{x},\Delta\mathbf{y})$. Hence, it's also independent of $(\boldsymbol{\delta},\boldsymbol{\theta})$. Moreover, since $J=J_1\sqcup J_2$ is an information set of $C$, the independent uniform vectors $(\mathbf{x},\mathbf{y})$ parametrize a uniformly random codeword $\mathbf{X}\in C$, whose $i$-th coordinate is $\ell_i(\mathbf{x},\mathbf{y})$ for every $i\in I$. Therefore, we obtain
    \begin{equation}\label{eq:high-support-rectangle-products}
    \mathcal{Q}= \mathbb{E}_{\boldsymbol{\delta},\boldsymbol{\theta}} \mathbb{E}_{\mathbf{X}\leftarrow C} \prod_{i\in I} q_{i,\boldsymbol{\delta},\boldsymbol{\theta}}(X_i).
    \end{equation}





    Let $C_I$ denote the projection of $C$ onto the coordinates in $I$. Since $\mathbf{X}$ is uniform on $C$, its projection $(X_i)_{i\in I}$ is uniform on $C_I$. By the properties of GRS codes, $C_I$ is a GRS code of length $m$ and dimension $\min\{m,t-1\}$. If $m\leq t-1$, then $C_I=\mathbb{F}_p^m$, so its coordinates are independent. Otherwise, $C_I$ has dimension $t-1$, together with $t-1\leq m\leq 11(t-1)/10,\ p>\max\{m,3000\}$, the parameters of $C_I$ satisfy the condition of Lemma~\ref{lem:decor}. Moreover, for fixed $\boldsymbol{\delta},\boldsymbol{\theta}$, each $q_{i,\boldsymbol{\delta},\boldsymbol{\theta}}$ is Boolean. By applying Lemma~\ref{lem:decor} to $C_I$, we obtain
	\begin{equation}
	 	\left| \mathbb{E}_{\mathbf{X}\leftarrow C}\prod_{i\in I}q_{i,\boldsymbol{\delta},\boldsymbol{\theta}}(X_i)- \prod_{i\in I}\mathbb{E}_{x\leftarrow\mathbb{F}_p} q_{i,\boldsymbol{\delta},\boldsymbol{\theta}}(x)\right| \le m^{O(1)}\gamma^{t-1}, 
		\notag
	 	\label{eq:high-support-decorrelation}
	  \end{equation}
    In fact, the left-hand side is zero when $m\le t-1$. Consequently, since $m\leq r$, for fixed $\boldsymbol{\delta},\boldsymbol{\theta}$, we have
    \[
    \mathbb{E}_{\mathbf{X}\leftarrow C}\prod_{i\in I}q_{i,\boldsymbol{\delta},\boldsymbol{\theta}}(X_i)=\prod_{i\in I}\mathbb{E}_{x\leftarrow\mathbb{F}_p} q_{i,\boldsymbol{\delta},\boldsymbol{\theta}}(x)+\varepsilon(\boldsymbol{\delta},\boldsymbol{\theta}),
    \]
	where $|\varepsilon(\boldsymbol{\delta},\boldsymbol{\theta})|\leq m^{O(1)}\gamma^{t-1}\leq r^{O(1)}\gamma^{t-1}$.
    
    For $i\in I$, define $K_i(s,z)=\mathbb{E}_{x\leftarrow\mathbb{F}_p} f_i(x)f_i(x+s)f_i(x+z)f_i(x+s+z)$. Then, we have $\mathbb{E}_{x\leftarrow\mathbb{F}_p}q_{i,\boldsymbol{\delta},\boldsymbol{\theta}}(x)=K_i(\delta_i,\theta_i).$ Combining \eqref{eq:high-support-rectangle-products} with the triangle inequality, we obtain
	\begin{equation}
	 	\mathcal{Q}=\mathbb{E}_{\boldsymbol{\delta},\boldsymbol{\theta}}\left(\prod_{i\in I}\mathbb{E}_{x\leftarrow\mathbb{F}_p} q_{i,\boldsymbol{\delta},\boldsymbol{\theta}}(x)+\varepsilon(\boldsymbol{\delta},\boldsymbol{\theta})\right)
	 	\leq\left|\mathbb E_{\boldsymbol{\delta},\boldsymbol{\theta}} \prod_{i\in I}K_i(\delta_i,\theta_i)\right|+r^{O(1)}\gamma^{t-1}.
	 	\label{eq:high-support-main-and-error}
	\end{equation}
	Let $S\subseteq I$ with $|S|=a$.  Since $|S\cup J_2|=a+b=t-1$, the MDS property of $C$ implies that the coordinate projection onto it is bijective. After the $J_2$ values are fixed, the increments $(\delta_i)_{i\in S}$ are consequently uniform on $\mathbb{F}_p^a$. Likewise, $S\cup J_1$ has $2a\le t-1$ coordinates, the corresponding projection has full rank, and after the $J_1$ values are fixed, $(\theta_i)_{i\in S}$ is uniform on $\mathbb{F}_p^a$. Since $\Delta \mathbf{x}$ and $\Delta \mathbf{y}$ are independent, the pairs $ Z_i:=(\delta_i,\theta_i)$ are jointly independent and uniform on $\mathbb{F}_p^2$.  Thus every $a$ of the $m$ variables $Z_i$ is jointly independent. Lemma~\ref{lem:limited-independence-holder} therefore gives
	\begin{align}
        \left|\mathbb{E}_{\boldsymbol{\delta},\boldsymbol{\theta}}\prod_{i\in I}K_i(\delta_i,\theta_i)\right|
        \le\prod_{i\in I}\|K_i\|_{m/a}.
	\label{eq:high-support-holder}
	\end{align}
    Since the $f_i$ is real-valued, expanding the square and writing $e=y-x$ also gives
	\begin{align*}
	    \|K_i\|_2^2 & =\mathbb E_{s,z,x,y}
	 	\prod_{\omega\in\{0,1\}^2} f_i(x+\omega_1s+\omega_2z)f_i(y+\omega_1s+\omega_2z)\\
	 	& =\mathbb E_{x,s,z,e}\prod_{\omega\in\{0,1\}^3}f_i(x+\omega_1s+\omega_2z+\omega_3e) =\|f_i\|_{U^3}^8.
	\end{align*}
	Hence, by $|\mathbb{E} [f_i]|\le\frac{7}{8}$ and Proposition \ref{prop:U^3 gowers}, we have
	\[
	\|K_i\|_2^2=\|f_i\|_{U^3}^8
	\le\frac{41}{100}=\eta.
	\]
	Moreover, $|K_i(s,t)|\le1$ since it is an average of products of $-1$ and $1$. Let $q:=m/a$, consider the following two cases:
	\begin{itemize}
	 	\item[a)] If $q\ge2$, then
	 	\[\|K_i\|_q^q=\mathbb{E}[|K_i|^{q-2}|K_i|^2]\le \|K_i\|_\infty^{q-2}\|K_i\|_2^2\le \|K_i\|_2^2\le\eta,\] and hence
	 	\[ \prod_{i\in I}\|K_i\|_q\le\eta^{m/q}=\eta^{a}.\]
	 	\item[b)] If $1\le q<2$, monotonicity of normalized $L^q$ norms gives $\|K_i\|_q\le\|K_i\|_2\le\eta^{1/2}$, so $\prod_{i\in I}\|K_i\|_q\leq \eta^{m/2}$.
	\end{itemize}
	Combining the two cases with \eqref{eq:high-support-main-and-error} and \eqref{eq:high-support-holder}, we obtain
	\[ \mathcal{Q} \le\eta^{\min\{m,2a\}/2}+r^{O(1)}\gamma^{t-1}.\]
	Finally, using \eqref{eq:prop3-four-point-cs} and $(x+y)^{1/4}\le x^{1/4}+y^{1/4}$ for $x,y\ge0$ give
	\[|\Gamma_{\mathbf{0}}(f_A)|
	\le\eta^{\frac{1}{8}\min\{m,2\lfloor(t-1)/2\rfloor\}}+r^{O(1)}\gamma^{(t-1)/4}.\]
	This completes the proof.
\end{proof}
	
\subsection{High Support Control with Large Bias}\label{sec:high-support-large-bias}
	
The following proposition controls the sum of $|\Gamma_{\mathbf{v}}(h_A)|$ over the high supports appearing in \eqref{eq:normalized-split}. As we shall see, the bound decreases as the average squared bias increases.
	
\begin{proposition}\label{prop:high-support-large-bias}
	Consider a GRS code $C\subseteq\mathbb F_p^n$ of dimension $t-1$, and let $f_1,\ldots,f_n:\mathbb F_p\to\{-1,1\}$ be leakage functions. Let $\mu_i$ and $h_i$ be as defined in Section \ref{sec:useful-results}. Suppose that $1/3<c=(t-1)/n<\frac12$ and $0<\varepsilon\le\frac12$. If $B<1$, then
	\begin{equation}
        \sum_{\substack{A\subseteq[n]\\|A|\geq(1-\varepsilon)n}}|\Gamma_{\mathbf{v}}(h_A)| \le\exp(nE_{\mathsf{high}}(c,\varepsilon,B)),
		\label{eq:high-support-large-bias}
	\end{equation}
	where $$E_{\mathsf{high}}(c,\varepsilon,B)=H(\varepsilon)+(1-\varepsilon)\left(c\log(1-B)+(1-2c)\log(1+B)\right).$$ The right-hand side is nonincreasing in $B\in[0,1)$. If $B=1$, all leakage functions are constant and all centred correlations in the sum vanish.
\end{proposition}
	
\begin{proof}
	Fix $A\subseteq[n]$ with $|A|=r\ge(1-\varepsilon)n$. Then $r\ge n/2\ge t-1$. Recall that any $t-1$ coordinates of a uniform word of $\mathbf{v}+C$ are jointly uniform and independent. Applying Lemma~\ref{lem:limited-independence-holder} with independence parameter $t-1$ to the $r$ coordinates in $A$, we obtain
	\[|\Gamma_{\mathbf{v}}(h_A)| \le\mathbb E_{\mathbf{X}\leftarrow\mathbf{v}+C} \prod_{i\in A}|h_i(X_i)| \le\prod_{i\in A}\|h_i\|_{r/(t-1)}.\]
	To use the same one-coordinate bound for every support, set $q=n/(t-1)=1/c\in(2,3)$. Since $r/(t-1)\le q$, monotonicity of normalized $L^q$ norms yields
	\[	|\Gamma_{\mathbf{v}}(h_A)|\le\prod_{i\in A}\|h_i\|_q.\]
	Since $h_i=f_i-\mu_i$ and $f_i$ is Boolean, we have $\mu_i=\Pr[f_i=1]-\Pr[f_i=-1]$. Hence, $|h_i|=1-\mu_i$, when $f_i=1$; $|h_i|=1+\mu_i$, when $f_i=-1$. This implies that
	\[\mathbb{E}\left[h_i^2\right]=1-\mu_i^2\quad \text{and} \quad \mathbb{E}\left[|h_i|^3\right]=(1-\mu_i^2)(1+\mu_i^2).\]
    For $2<q<3$, the numbers $3-q$ and $q-2$ are nonnegative and sum to one. Hence, $|h_i|^q =(|h_i|^2)^{3-q}(|h_i|^3)^{q-2}$. Then the H\"older's inequality with conjugate exponents $1/(3-q)$ and $1/(q-2)$ yields
	\[
    \mathbb{E}\left[|h_i|^q\right] \le(\mathbb{E}\left[h_i^2\right])^{3-q}(\mathbb{E}\left[|h_i|^3\right])^{q-2}.
	\]
	Taking the $q$th root and using $q=1/c$, we obtain $\|h_i\|_q\le(1-\mu_i^2)^c(1+\mu_i^2)^{1-2c}$. The same bound holds when $f_i$ is constant, since then $h_i=0$.
		
	Define $\phi_c(x)=(1-x)^c(1+x)^{1-2c},~0\le x\le1$. In the above analysis, we have proved
	\[|\Gamma_{\mathbf{v}}(h_A)|\le\prod_{i\in A}\phi_c(\mu_i^2).\]
	whenever $|A|\geq (1-\varepsilon)n$. Note that direct differentiation yields
	\begin{align*}
		\phi_c'(x) & =\phi_c(x)\frac{1-3c-(1-c)x}{1-x^2};\\
		\phi_c''(x) & =\phi_c(x)\frac{c(9c-5+2(1-3c)x-(1-c)x^2)}{(1-x^2)^2}.
	\end{align*}
	Since $1/3< c< 1/2$, we have both $\phi_c'(x)<0$ and $\phi_c''(x)<0$ for $x\in (0,1)$. Hence, $\phi_c(x)$ is concave and nonincreasing with $\phi_c(0)=1$ and $\phi_c(1)=0$ on $[0,1]$. Therefore, we have $0\le \phi_c(x)\le1$ for $x\in [0,1]$ and $\phi_c(B)>0$ when $B<1$.
		
	Define
	\[S=\sum_{\substack{A\subseteq[n]\\|A|\ge(1-\varepsilon)n}}\prod_{i\in A}\phi_c(\mu_i^2).\]
	To estimate this sum, introduce an auxiliary parameter $s\in(0,1]$. For every $A\subseteq[n]$ with $|A|\geq(1-\varepsilon)n$, we have $s^{n-|A|}\geq s^{\varepsilon n}$. Hence,
	\begin{align*}
		S&\le s^{-\varepsilon n}s^{n-|A|} \sum_{\substack{A\subseteq[n]\\|A|\ge(1-\varepsilon)n}}\prod_{i\in A}\phi_c(\mu_i^2) 
		\le s^{-\varepsilon n} \sum_{A\subseteq[n]} s^{n-|A|}\prod_{i\in A}\phi_c(\mu_i^2) 
		= s^{-\varepsilon n} \prod_{i=1}^n\bigl(\phi_c(\mu_i^2)+s\bigr).
	\end{align*}
	For fixed $s>0$, since $\phi_c(x)$ is concave and nonnegative on $[0,1]$, the function $\ln(\phi_c(x)+s)$ is also concave. Therefore, by Jensen's inequality,
	\[
    \prod_{i=1}^n\bigl(\phi_c(\mu_i^2)+s\bigr) =\exp\left(\sum_{i=1}^{n}\ln(\phi_c(\mu_i^2)+s)\right) \le\exp\left(n\ln\left(\phi_c(\frac{1}{n}\sum_{i=1}^{n}\mu_i^2)+s\right)\right)
	=\bigl(\phi_c(B)+s\bigr)^n.
	\]
	Consequently, for $s\in (0,1]$, it holds that $S\leq s^{-\varepsilon n}\bigl(\phi_c(B)+s\bigr)^n$. Optimizing this bound with respect to $s$, we choose $s=\frac{\varepsilon\phi_c(B)}{1-\varepsilon}\in(0,1]$. Substituting into the preceding bounds gives
	\begin{align*}
		S \le\left(\frac{\varepsilon\phi_c(B)}{1-\varepsilon}\right)^{-\varepsilon n}
		\left(\frac{\phi_c(B)}{1-\varepsilon}\right)^n=\exp\left(nE_{\mathsf{high}}(c,\varepsilon,B)\right).
	\end{align*}
	This proves~\eqref{eq:high-support-large-bias}. The monotonicity in $B$ follows from that of $\phi_c$. Finally, if $B=1$, then $\mu_i^2=1$ for every $i$. Hence, all $f_i$ are constant leakage functions and $h_i=0$, so $\Gamma_{\mathbf{v}}(h_A)=0$ for every nonempty $A$.
\end{proof}
	
\section{Proof of the Main Theorem}\label{sec:main-theorem-proof}
	
In this section, we use Theorem~\ref{thm:output-decomposition} to combine the low and high support bounds established in  Sections~\ref{sec:low-support} and~\ref{sec:high-support}. With a suitable choice of parameters, we obtain the following main result.
	
\begin{theorem}[Main theorem]\label{thm:main-leakage-resilience}
	There is an absolute constant $n_0$ such that following holds for every $n\ge n_0$ and every prime $p>n$. Consider an $(n,t)$-Shamir's secret sharing scheme over $\mathbb{F}_p$. If the reconstruction  threshold satisfies $t\ge 0.489n$. Then, for any $1$-bit leakage functions $f_i:\mathbb{F}_p\to\{-1,1\}$, and any two secrets $s_1,s_2\in\mathbb{F}_p$,
	\[\mathsf{SD}\!\left(\mathbf{f}(\mathsf{share}(s_1)),\mathbf{f}(\mathsf{share}(s_2))\right)\le\exp(-\Omega(n)).\]
\end{theorem}
	
\begin{proof}
	By the threshold monotonicity of Shamir's scheme, it suffices to consider $t=\lceil489n/1000\rceil$. Let $C=\mathsf{GRS}_{\boldsymbol{\alpha},\boldsymbol{\alpha}}(n,t-1)$ and consider the affine shifts $\mathbf{v}=s_i\mathbf{1},\mathbf{w}=s_2\mathbf{1}$. We choose
	\[
	t=\left\lceil\frac{489n}{1000}\right\rceil,\qquad
	R=t-1-\left\lceil\frac{n}{10000}\right\rceil,\qquad
	B_0=\frac{21}{100},\qquad u=\frac65.
	\] 
    and recall
	\[a=\frac{R}{n},\qquad c=\frac{t-1}{n} ,\qquad r_0=t-1+R,\qquad \varepsilon=1-\frac{r_0}{n}.\]
	As $n\to\infty$, these parameters satisfy $c\to0.489,\ a\to0.4889,\ \varepsilon\to0.0221$. If $B=1$, the leakage functions are constant, so we assume $B<1$.
		
	\begin{itemize}
		\item[a).] For $B\ge B_0$, apply \eqref{eq:normalized-split} with Lemma~\ref{lem:normalized-low-support} and Proposition \ref{prop:high-support-large-bias}, together with the monotonicity of the latter bound in $B$ gives
		\begin{equation}
			\mathsf{SD}(\mathbf{f}(\mathsf{share}(s_1)),\mathbf{f}(\mathsf{share}(s_2))) \le n^{O(1)}\exp\left(\frac{n}{2}E_{\mathsf{low}}(c,a)\right) +\exp\Bigl(nE_{\mathsf{high}}(c,\varepsilon,B_0)\Bigr).
			\label{eq:main-large-bias-distance}
		\end{equation}
			
		\item[b).] For $B\le B_0$, applying Lemmas \ref{lem:normalized-low-support-transfer} and \ref{lem:normalized-low-support} to the low support term in \eqref{eq:walsh-split}, we obtain 
        $$
        \frac{1}{2}\left(\sum_{t\leq|A|\leq r_0}|a_A^f|^2\right)^{1/2}\leq n^{O(1)}u^{-t}\lambda_{u^2}(B_0)^{n/2}\exp\left(\frac{n}{2}E_{\mathsf{low}}(c,a)\right).
        $$ 
        Since $\sum_{i\in [n]}\mu_i^2=nB\leq nB_0$, the number of coordinates satisfying $|\mu_i|>7/8$ is at most $nB_0/(7/8)^2<t-1$. For any $A\subseteq[n]$ with $|A|>r_0$, choose $J\subseteq A$ of size $t-1$ containing all such coordinates in $A$. Setting $I=A\setminus J$, we have $ R+1\le |I|:=m\le n-t+1$. For sufficiently large $n$, our choice of parameters ensures $\left\lceil\frac{t-1}{2}\right\rceil\le R\le t-2, n-t+1\le\frac{11}{10}(t-1)$. Thus, Proposition~\ref{prop:large-support-small-bias} gives $\eta^{R/8}+n^{O(1)}\gamma^{(t-1)/4}$. There are at most $n^{O(1)}e^{nH(\varepsilon)}$ high supports. Summing squares and taking the square root therefore gives
		\begin{align}
			&\mathsf{SD}(\mathbf{f}(\mathsf{share}(s_1)),\mathbf{f}(\mathsf{share}(s_2)))\\ 
            \leq & n^{O(1)}\Bigl(u^{-t}\lambda_{u^2}(B_0)^{n/2}\exp\left(\frac{n}{2}E_{\mathsf{low}}(c,a)\right)+\exp\left(\frac{n}{2}H(\varepsilon)\right)\bigl(\eta^{R/8}+\gamma^{(t-1)/4}\bigr)\Bigr).
            \label{eq:main-small-bias-distance}
		\end{align}
		\end{itemize}
	The limiting exponential rates in \eqref{eq:main-large-bias-distance} and \eqref{eq:main-small-bias-distance} are as follows:
	\begin{center}
    \begin{tabular}{ccc}
    \toprule
        Case & Low support & High support \\
        \midrule
        $B\geq B_0$ & $-0.173229$ & $-0.002517$ \\
        $B<B_0$ & $-0.007800$ & $-0.001436,\ -0.001507$ \\
        \bottomrule
    \end{tabular}
    \end{center}
	They are all strictly negative. Hence, we have \[\mathsf{SD}(\mathbf{f}(\mathsf{share}(s_1)),\mathbf{f}(\mathsf{share}(s_2)))\le\exp(-\Omega(n)).\]
    This completes the proof.
\end{proof}
	
\section*{Acknowledgment}
The authors used ChatGPT (GPT-6 Astra and GPT-6.1 Sol) as interactive research assistants during this work. The models were used to discuss and refine mathematical ideas, examine proof strategies, check calculations and parameter choices, identify relevant literature, and assist with drafting and revising the manuscript. In particular, discussions with the models helped explore techniques involving Fourier analysis and list recovery of Reed--Solomon codes. The authors independently evaluated and verified the mathematical suggestions before incorporating them into the work. The authors developed the main research ideas and take full responsibility for the correctness of the results, proofs, references, and final manuscript.

\bibliographystyle{IEEEtranS}
\bibliography{reference}

@article{10.1145/359168.359176,
  author={Shamir, Adi},
  title={How to Share a Secret},
  journal={Communications of the ACM},
  volume={22},
  number={11},
  pages={612--613},
  year={1979},
  publisher={Association for Computing Machinery},
  doi={10.1145/359168.359176}
}

@inproceedings{blakley1979safeguarding,
  author={Blakley, George Robert},
  title={Safeguarding Cryptographic Keys},
  booktitle={Proceedings of the 1979 AFIPS National Computer Conference},
  volume={48},
  pages={313--317},
  year={1979},
  publisher={AFIPS Press},
  doi={10.1109/AFIPS.1979.98}
}

@inproceedings{10.5555/646761.706156, 
  author={Kocher, Paul C.},
  title={Timing Attacks on Implementations of {Diffie--Hellman}, {RSA}, {DSS}, and Other Systems},
  booktitle={Advances in Cryptology---CRYPTO 1996},
  series={Lecture Notes in Computer Science},
  volume={1109},
  pages={104--113},
  year={1996},
  publisher={Springer},
  doi={10.1007/3-540-68697-5_9}
}

@inproceedings{kocher1999differential,
  author={Kocher, Paul C. and Jaffe, Joshua and Jun, Benjamin},
  title={Differential Power Analysis},
  booktitle={Advances in Cryptology---CRYPTO 1999},
  series={Lecture Notes in Computer Science},
  volume={1666},
  pages={388--397},
  year={1999},
  publisher={Springer},
  doi={10.1007/3-540-48405-1_25}
}

@inproceedings{benhamouda2018local,
  author={Benhamouda, Fabrice and Degwekar, Akshay and Ishai, Yuval and Rabin, Tal},
  title={On the Local Leakage Resilience of Linear Secret Sharing Schemes},
  booktitle={Advances in Cryptology---CRYPTO 2018},
  pages={531--561},
  year={2018},
  publisher={Springer},
  doi={10.1007/978-3-319-96884-1_18}
}

@article{benhamouda2021local,
  author={Benhamouda, Fabrice and Degwekar, Akshay and Ishai, Yuval and Rabin, Tal},
  title={On the Local Leakage Resilience of Linear Secret Sharing Schemes},
  journal={Journal of Cryptology},
  volume={34},
  number={2},
  pages={10},
  year={2021},
  publisher={Springer},
  doi={10.1007/s00145-021-09375-2}
}

@inproceedings{goyal2018non,
  author={Goyal, Vipul and Kumar, Ashutosh},
  title={Non-Malleable Secret Sharing},
  booktitle={Proceedings of the 50th Annual ACM SIGACT Symposium on Theory of Computing (STOC 2018)},
  pages={685--698},
  year={2018},
  publisher={Association for Computing Machinery},
  doi={10.1145/3188745.3188872}
}

@inproceedings{ishai2003private,
  author={Ishai, Yuval and Sahai, Amit and Wagner, David},
  title={Private Circuits: Securing Hardware against Probing Attacks},
  booktitle={Advances in Cryptology---CRYPTO 2003},
  series={Lecture Notes in Computer Science},
  volume={2729},
  pages={463--481},
  year={2003},
  publisher={Springer},
  doi={10.1007/978-3-540-45146-4_27}
}

@inproceedings{maji2021constructing,
  author={Maji, Hemanta K. and Paskin-Cherniavsky, Anat and Suad, Tom and Wang, Mingyuan},
  title={Constructing Locally Leakage-Resilient Linear Secret-Sharing Schemes},
  booktitle={Advances in Cryptology---CRYPTO 2021},
  pages={779--808},
  year={2021},
  publisher={Springer},
  doi={10.1007/978-3-030-84252-9_26}
}

@inproceedings{maji2022improved,
  author={Maji, Hemanta K. and Nguyen, Hai H. and Paskin-Cherniavsky, Anat and Wang, Mingyuan},
  title={Improved Bound on the Local Leakage-Resilience of Shamir's Secret Sharing},
  booktitle={2022 IEEE International Symposium on Information Theory (ISIT 2022)},
  pages={2678--2683},
  year={2022},
  publisher={IEEE},
  doi={10.1109/ISIT50566.2022.9834695}
}

@inproceedings{klein2023new,
  author={Klein, Ohad and Komargodski, Ilan},
  title={New Bounds on the Local Leakage Resilience of Shamir's Secret Sharing Scheme},
  booktitle={Advances in Cryptology---CRYPTO 2023},
  pages={139--170},
  year={2023},
  publisher={Springer},
  doi={10.1007/978-3-031-38557-5_5}
}

@inproceedings{Kasser2024Improvement,
  author={Kasser, Dustin},
  title={An Improvement Upon the Bounds for the Local Leakage Resilience of Shamir's Secret Sharing Scheme},
  booktitle={Theory of Cryptography---TCC 2024},
  pages={395--422},
  year={2024},
  publisher={Springer},
  doi={10.1007/978-3-031-78023-3_13}
}

@inproceedings{nguyen2024towards,
  author={Nguyen, Hai H.},
  title={Towards Breaking the Half-Barrier of Local Leakage-Resilient Shamir's Secret Sharing},
  booktitle={Advances in Cryptology---CRYPTO 2024},
  pages={257--285},
  year={2024},
  publisher={Springer},
  doi={10.1007/978-3-031-68388-6_10}
}

@inproceedings{maji2021leakage,
  author={Maji, Hemanta K. and Nguyen, Hai H. and Paskin-Cherniavsky, Anat and Suad, Tom and Wang, Mingyuan},
  title={Leakage-Resilience of the Shamir Secret-Sharing Scheme against Physical-Bit Leakages},
  booktitle={Advances in Cryptology---EUROCRYPT 2021},
  pages={344--374},
  year={2021},
  publisher={Springer},
  doi={10.1007/978-3-030-77886-6_12}
}

@inproceedings{nielsen2020lower,
  author={Nielsen, Jesper Buus and Simkin, Mark},
  title={Lower Bounds for Leakage-Resilient Secret Sharing},
  booktitle={Advances in Cryptology---EUROCRYPT 2020},
  pages={556--577},
  year={2020},
  publisher={Springer},
  doi={10.1007/978-3-030-45721-1_20}
}

@inproceedings{maji2024constructing,
  author={Maji, Hemanta K. and Nguyen, Hai H. and Paskin-Cherniavsky, Anat and Ye, Xiuyu},
  title={Constructing Leakage-Resilient Shamir's Secret Sharing: Over Composite Order Fields},
  booktitle={Advances in Cryptology---EUROCRYPT 2024},
  pages={286--315},
  year={2024},
  publisher={Springer},
  doi={10.1007/978-3-031-58737-5_11}
}

@inproceedings{nguyen2025physical,
  author={Nguyen, Hai H.},
  title={Physical-Bit Leakage Resilience of Linear Code-Based Secret Sharing},
  booktitle={Advances in Cryptology---EUROCRYPT 2025},
  pages={64--93},
  year={2025},
  publisher={Springer},
  doi={10.1007/978-3-031-91101-9_3}
}

@inproceedings{guruswami2016repairing,
  author={Guruswami, Venkatesan and Wootters, Mary},
  title={Repairing {Reed--Solomon} Codes},
  booktitle={Proceedings of the 48th Annual ACM Symposium on Theory of Computing (STOC 2016)},
  pages={216--226},
  year={2016},
  publisher={Association for Computing Machinery},
  doi={10.1145/2897518.2897525}
}

@article{guruswami2017repairing,
  author={Guruswami, Venkatesan and Wootters, Mary},
  title={Repairing {Reed--Solomon} Codes},
  journal={IEEE Transactions on Information Theory},
  volume={63},
  number={9},
  pages={5684--5698},
  year={2017},
  publisher={IEEE},
  doi={10.1109/TIT.2017.2702660}
}

@article{con2022nonlinear,
  author={Con, Roni and Tamo, Itzhak},
  title={Nonlinear Repair of Reed--Solomon Codes},
  journal={IEEE Transactions on Information Theory},
  volume={68},
  number={8},
  pages={5165--5177},
  year={2022},
  publisher={IEEE},
  doi={10.1109/TIT.2022.3167615}
}

@inproceedings{con2023repairing,
  author={Con, Roni and Shutty, Noah and Tamo, Itzhak
	and Wootters, Mary},
  title={Repairing {Reed--Solomon} Codes over Prime Fields via Exponential Sums},
  booktitle={2023 IEEE International Symposium on Information Theory (ISIT 2023)},
  pages={1330--1335},
  year={2023},
  publisher={IEEE},
  doi={10.1109/ISIT54713.2023.10206937}
}

@misc{cryptoeprint:2026/833,
  author={Boudgoust, Katharina and Simkin, Mark},
  title={Scale, Round, Break: Simple Leakage Attacks on Secret Sharing Schemes},
  howpublished={Cryptology ePrint Archive, Paper 2026/833},
  year={2026},
  url={https://eprint.iacr.org/2026/833}
}

@book{Tao_Vu_2006,
  author    = {Terence Tao and Van H. Vu},
  title     = {Additive Combinatorics},
  series    = {Cambridge Studies in Advanced Mathematics},
  volume    = {105},
  publisher = {Cambridge University Press},
  year      = {2006},
  isbn      = {978-0-521-85386-6},
  doi       = {10.1017/CBO9780511755149}
}

@book{HuffmanPless2003,
  author    = {W. Cary Huffman and Vera Pless},
  title     = {Fundamentals of Error-Correcting Codes},
  publisher = {Cambridge University Press},
  year      = {2003},
  doi       = {10.1017/CBO9780511807077},
  isbn      = {9780511807077}
}

@misc{SunWootters2026,
	author        = {Yihang Sun and Mary Wootters},
	title         = {Improved Local Leakage Resilience of Shamir Secret Sharing and Worst-Case Optimal Polynomial Intersection},
	year          = {2026},
	eprint        = {2610.12357},
	archivePrefix = {arXiv},
	primaryClass  = {quant-ph},
	url           = {https://arxiv.org/abs/2610.12357}
}

@misc{Nguyen2026,
	author        = {Hai H. Nguyen},
	title         = {Local Leakage Resilience of Shamir Secret Sharing: Below Half Threshold Rate},
	year          = {2026},
	eprint        = {2610.17711},
	archivePrefix = {arXiv},
	primaryClass  = {cs.CR},
	url           = {https://arxiv.org/abs/2610.17711}
}

@article{Pollard1974,
  author  = {Pollard, J. M.},
  title   = {A Generalisation of the Theorem of Cauchy and Davenport},
  journal = {Journal of the London Mathematical Society},
  series  = {2},
  volume  = {8},
  number  = {3},
  pages   = {460--462},
  year    = {1974},
  doi     = {10.1112/jlms/s2-8.3.460}
}

@misc{Jeronimo2026,
  author        = {Fernando Granha Jeronimo},
  title         = {Algorithmic List Decoding at Capacity and
                   Optimal Proximity Gaps for Reed-Solomon Codes},
  year          = {2026},
  eprint        = {2609.05870},
  archivePrefix = {arXiv},
  primaryClass  = {cs.IT},
  doi           = {10.48550/arXiv.2609.05870},
  url           = {https://arxiv.org/abs/2609.05870}
}

@inproceedings{BHMY26,
  author    = {Aniruddha Biswas and Jihun Hwang and Hemanta K. Maji and Xiuyu Ye},
  title     = {Resilience of Inner-Product Masking Scheme Against Hamming Weight Leakage},
  booktitle = {7th Conference on Information-Theoretic Cryptography (ITC 2026)},
  series    = {LIPIcs},
  volume    = {385},
  pages     = {9:1--9:23},
  year      = {2026},
  publisher = {Schloss Dagstuhl},
  doi       = {10.4230/LIPIcs.ITC.2026.9}
}

@misc{BHMSY26,
  author = {Aniruddha Biswas and Jihun Hwang and Hemanta K. Maji and Ilya D. Shkredov and Xiuyu Ye},
  title  = {Beyond Threshold Security: Additive Secret Sharing under Hamming-Weight Leakage},
  year   = {2026},
  note   = {Manuscript},
  url    = {https://www.cs.purdue.edu/homes/hmaji/papers/BHMSY26.pdf}
}

@inproceedings{HS23,
  author    = {Charlotte Hoffmann and Mark Simkin},
  title     = {Stronger Lower Bounds for Leakage-Resilient Secret Sharing},
  booktitle = {Progress in Cryptology -- LATINCRYPT 2023},
  series    = {Lecture Notes in Computer Science},
  volume    = {14168},
  pages     = {215--228},
  publisher = {Springer},
  year      = {2023},
  doi       = {10.1007/978-3-031-44469-2_11}
}

@article{Sud97,
  author  = {Madhu Sudan},
  title   = {Decoding of Reed Solomon Codes beyond the Error-Correction Bound},
  journal = {Journal of Complexity},
  volume  = {13},
  number  = {1},
  pages   = {180--193},
  year    = {1997},
  doi     = {10.1006/jcom.1997.0439}
}

@article{GS99,
  author  = {Venkatesan Guruswami and Madhu Sudan},
  title   = {Improved Decoding of Reed-Solomon and Algebraic-Geometry Codes},
  journal = {IEEE Transactions on Information Theory},
  volume  = {45},
  number  = {6},
  pages   = {1757--1767},
  year    = {1999},
  doi     = {10.1109/18.782097}
}

@article{BGM23,
  author  = {Joshua Brakensiek and Sivakanth Gopi and Visu Makam},
  title   = {Generic Reed-Solomon Codes Achieve List-Decoding Capacity},
  journal = {SIAM Journal on Computing},
  year    = {2024},
  doi     = {10.1137/23M1598064},
  note    = {Conference version in STOC 2023}
}

@article{AGG25,
  author  = {Omar Alrabiah and Zeyu Guo and Venkatesan Guruswami and Ray Li and Zihan Zhang},
  title   = {Random Reed-Solomon Codes Achieve List-Decoding Capacity With Linear-Sized Alphabets},
  journal = {Advances in Combinatorics},
  volume  = {2025},
  pages   = {8},
  year    = {2025},
  doi     = {10.19086/aic.2025.8}
}

@misc{BCP26,
  author        = {Joshua Brakensiek and Yeyuan Chen and Aaron Putterman and Zihan Zhang and Kai Zhe Zheng},
  title         = {Algorithmic List Decoding of Reed-Solomon Codes up to Capacity},
  year          = {2026},
  eprint        = {2609.08005},
  archivePrefix = {arXiv},
  primaryClass  = {cs.IT},
  url           = {https://arxiv.org/abs/2609.08005}
}

@inproceedings{RW14,
  author    = {Atri Rudra and Mary Wootters},
  title     = {Every List-Decodable Code for High Noise Has Abundant Near-Optimal Rate Puncturings},
  booktitle = {Proceedings of the 46th Annual ACM Symposium on Theory of Computing},
  pages     = {764--773},
  publisher = {ACM},
  year      = {2014},
  doi       = {10.1145/2591796.2591797}
}

@inproceedings{ST20,
  author    = {Chong Shangguan and Itzhak Tamo},
  title     = {Combinatorial List-Decoding of Reed-Solomon Codes Beyond the Johnson Radius},
  booktitle = {Proceedings of the 52nd Annual ACM SIGACT Symposium on Theory of Computing},
  pages     = {538--551},
  publisher = {ACM},
  year      = {2020},
  doi       = {10.1145/3357713.3384295}
}

@article{GLSTW24,
  author  = {Zeyu Guo and Ray Li and Chong Shangguan and Itzhak Tamo and Mary Wootters},
  title   = {Improved List-Decodability and List-Recoverability of Reed-Solomon Codes via Tree Packings},
  journal = {SIAM Journal on Computing},
  volume  = {53},
  number  = {2},
  pages   = {389--430},
  year    = {2024},
  doi     = {10.1137/21M1463707}
}

@misc{DKT26,
  author       = {Quang Dao and Scott Duke Kominers and Justin Thaler},
  title        = {Reed-Solomon Codes Beyond Johnson: Efficient Decoding and Smaller Cryptographic Proofs},
  howpublished = {Cryptology ePrint Archive, Paper 2026/2056},
  year         = {2026},
  url          = {https://eprint.iacr.org/2026/2056}
}

@unpublished{HMSY26,
  author = {Jihun Hwang and Hemanta K. Maji and Ilya D. Shkredov and Xiuyu Ye},
  title  = {Local Leakage Resilience of Multiplication-Friendly Shamir Secret Sharing},
  note   = {Manuscript, October 28, 2026},
  year   = {2026}
}

@book{Tao2012HigherOrder,
  author    = {Terence Tao},
  title     = {Higher Order Fourier Analysis},
  series    = {Graduate Studies in Mathematics},
  volume    = {142},
  publisher = {American Mathematical Society},
  year      = {2012},
  isbn      = {9780821889862}
}

@misc{rong-github,
  author       = {Zhang, Cheng and Wang, Xin and Kong, Xiangliang},
  title        = {Improved Threshold Bound for the Local Leakage Resilience of Shamir's Secret Sharing Scheme},
  howpublished = {Available at GitHub: https://github.com/rongxlkong-jpg/Local-Leakage-Resilience-of-Shamir-s-Secret-Sharing-Scheme-/blob/main/Improved\_Threshold\_\_Bound\_for\_the\_Local\_Leakage\_Resilience\_of\_Shamir\_s\_Secret\_Sharing\_Scheme.pdf}}

\appendices
\section{Proof of the Gowers \texorpdfstring{$U^3$}{U3} Bound}\label{app:gowers}

We first recall Pollard's theorem and apply it to derive an auxiliary $U^2$ estimate, which will be used to prove Proposition \ref{prop:U^3 gowers}.
	
\begin{theorem}[{\cite[Pollard's theorem]{Pollard1974}}]
\label{thm:pollard}
	Let $p$ be a prime, and let $A,B\subseteq\mathbb{F}_p$ be nonempty subsets, and let $1\le j\le\min\{|A|,|B|\}$ be an integer. For $x\in\mathbb{F}_p$, define $ r_{A+B}(x) =\bigl|\{(a,b)\in A\times B:a+b=x\}\bigr|.$ Then
	\[\sum_{x\in\mathbb{F}_p} \min\{j,r_{A+B}(x)\} \ge j\min\{p,|A|+|B|-j\}.\]
\end{theorem}
	
\begin{lemma}
\label{lem:bias-U2-bound}
	Let $ f : \mathbb{F}_p  \to \{-1,1\} $ be a leakage function, and define $ \beta=|\mathbb{E} f| $.  Then
	\[ \|f\|_{U^2}^4 \leq F(\beta)+\frac{8}{3p^2}, \quad
	\text{where} \quad F(\theta)=\frac{1}{3}-2\theta^2+\frac{8}{3}|\theta|^3.\]
\end{lemma}
\begin{proof}
	Since replacing $f$ by $-f$ does not change either $\|f\|_{U^2}$ or $\beta$, we may assume that $\mathbb{E} f\leq 0$. Let $S=\{x\in\mathbb {F}_p: f(x)=1\}$, put $m=|S|$ and define $\rho=m/p$.  Then $m\leq p/2$, $f=2\mathbbm{1}_S-1$, and $\beta=-\mathbb{E}f=1-2\rho, \rho=(1-\beta)/2$. If $m=0$, then $f \equiv -1$, so $\beta=1$ and $\|f\|_{U^2}^4=\sum_a|\widehat{f}(a)|^4=1=F(1)$. Thus the claim follows. 
		
	Hence, assume $m \geq 1$. For $x \in \mathbb{F}_p$, write $r(x)=|\{(u,v) \in S^2: u+v=x\}|$. Since $m\leq p/2$, Theorem \ref{thm:pollard} gives, for every integer $ 1\leq j \leq m$, 
	\begin{equation}
		\sum_{x\in\mathbb{F}_p} \min\{j,r(x)\} \geq j\min\{p, 2m-j\}=j(2m-j). \label{eq: Pollard's inequality}
	\end{equation}
	Since every ordered pair in $S^2$ contributes to exactly one value of $x$, we have $\sum_x r(x)=m^2$. Moreover, for every integer $0\leq u \leq m $,
	\[\sum_{j=1}^m\min\{u,j\}=\sum_{j=1}^{u}j+\sum_{j=u+1}^{m}u
	=\frac{u(u+1)}2+(m-u)u
	=\frac{(2m+1)u-u^2}{2}. \]
	Applying this identity with $u=r(x)$ and using \eqref{eq: Pollard's inequality}, we obtain
	\[
	\begin{aligned}
		\sum_x r(x)^2
		&=(2m+1)\sum_{x}r(x)
        -2\sum_{j=1}^m\sum_x\min\{j,r(x)\} \\
		&\leq (2m+1)m^2
		-2\sum_{j=1}^m j(2m-j)
		=\frac{2m^3+m}{3}.
	\end{aligned}
	\]
	By the definition of the Fourier transform and the orthogonality of additive characters, it follows that
	\[
	\begin{aligned}
		\sum_{a\in\mathbb {F}_p}|\widehat{\mathbbm{1}_S}(a)|^4
		&=\sum_{a\in\mathbb{F}_p}
		\left|\frac{1}{p}\sum_{x\in S}\overline{\chi_{a}(x)}\right|^4\\
		&=\frac1{p^4}
		\sum_{u_1,v_1,u_2,v_2\in S}
		\sum_{a\in\mathbb{F}_p}
		e_p\big(-a(u_1+v_1-u_2-v_2)\big)\\
		&=\frac{p}{p^4}\left|
		\left\{(u_1,v_1,u_2,v_2)\in S^4: u_1+v_1=u_2+v_2\right\} \right|\\
		&=\frac{1}{p^3}\sum_{x\in\mathbb{F}_p}r(x)^2
		\leq\frac{2m^3+m}{3p^3}.
	\end{aligned}
	\]
	Since $\widehat{\mathbbm{1}_S}(0)=\rho$ and $f=2\mathbbm{1}_S-1$, the Fourier coefficients of $f$ satisfy $\widehat{f}(0)=2\rho-1=-\beta, \widehat{f}(a)=2\widehat{\mathbbm{1}_S}(a),
	a\neq0$. Consequently,
	\[
	\begin{aligned}
		\|f\|_{U^2}^4
		&=\sum_{a\in\mathbb{F}_p}|\widehat{f}(a)|^4
		=|\widehat{f}(0)|^4+\sum_{a\neq0}|\widehat{f}(a)|^4\\
		&=\beta^4+16\sum_{a\neq0}|\widehat{\mathbbm{1}_S}(a)|^4
		=\beta^4+16\left(\sum_{a\in\mathbb{F}_p}|\widehat{\mathbbm{1}_S}(a)|^4-\rho^4\right)\\
		&\leq\beta^4+16\left(
		\frac{2m^3+m}{3p^3}-\rho^4\right)
		=\beta^4+16\left(\frac{2}{3}\rho^3-\rho^4\right)
		+\frac{16\rho}{3p^2}\\
		&=F(\beta)+\frac{16\rho}{3p^2}\leq F(\beta)+\frac8{3p^2},
	\end{aligned}
	\]
	where the last equality follows from $\rho=(1-\beta)/2$, and the last inequality uses $\rho\le1/2$. This completes the proof. 
\end{proof}
	
\begin{proof}[Proof of Proposition \ref{prop:U^3 gowers}]
	Define $K(z)=\mathbb E_{x\leftarrow\mathbb{F}_p} f(x)f(x+z)$. By the recursive identity for Gowers norms, $ \|f\|_{U^3}^8=\mathbb{E}_{z\leftarrow\mathbb{F}_p}
	\|f(x)f(x+z)\|_{U^2}^4$. Since $f(x)f(x+z)$ is Boolean with mean $K(z)$, applying Lemma \ref{lem:bias-U2-bound} and using the fact that $F$ is even yields
	\begin{equation}
		\|f\|_{U^3}^8
		\leq\mathbb{E}_{z\leftarrow \mathbb{F}_p} F(K(z))+\frac{8}{3p^2}. \label{eq:gowers-u3-autocorrelation}
	\end{equation}
	To bound the right-hand side, we introduce the function
	\[G(\theta)=\max\left\{\frac{1}{3},F(\theta)\right\}=
	\begin{cases}
		\frac{1}{3},&|\theta|\leq\frac{3}{4},\\[4pt]
		F(\theta),&\frac{3}{4}<|\theta|\le1.
	\end{cases}\]
	Recall that $F(\theta)=\frac13-2\theta^2+\frac83|\theta|^3$. Since $F''(\theta)=-4+16|\theta|>0$ for $3/4<|\theta|\leq1$, the function $G$ is convex on each of its three defining intervals. Moreover, the one-sided derivatives of $G$ increase at $\theta=\pm3/4$. Hence $G$ is convex on $[-1,1]$. Furthermore, since $|G'(\theta)|\leq4$ wherever it is differentiable, $G$ is $4$-Lipschitz. As $F\leq G$,  \eqref{eq:gowers-u3-autocorrelation} implies that $\|f\|_{U^3}^8\le\mathbb {E}_{z\leftarrow \mathbb{F}_p}G(K(z))+8/(3p^2)$.
		
	As in the proof of Lemma \ref{lem:bias-U2-bound}, we may assume without loss of generality that $\mathbb{E} f\leq0$, write $f=2\mathbbm{1}_S-1$, where $m=|S|\leq p/2$, $\rho=m/p$, and $\beta=|\mathbb{E}f|=1-2\rho$. For $z\in\mathbb {F}_p$, let $r_{S-S}(z)=|\{(u,v)\in S^2: u-v=z\}|$. We can now express $K(z)$ in terms of $r_{S-S}(z)$: 
	\[
	\begin{aligned}
		K(z)
		&=\mathbb{E}_{x\leftarrow\mathbb{F}_p} f(x)f(x+z)
		=\mathbb {E}_{x\leftarrow\mathbb{F}_p}\big[(2\mathbbm{1}_S(x)-1)(2\mathbbm{1}_S(x+z)-1)\big]\\
		&=4\mathbb {E}_{x\leftarrow\mathbb{F}_p}[\mathbbm{1}_S(x)\mathbbm{1}_S(x+z)]-2\mathbb{E}_{x\leftarrow\mathbb{F}_p}\mathbbm{1}_S(x)-2\mathbb{E}_{x\leftarrow\mathbb{F}_p}\mathbbm{1}_S(x+z)+1\\
		&=4\mathbb {E}_{x\leftarrow\mathbb{F}_p}[\mathbbm{1}_S(x)\mathbbm{1}_S(x+z)]-4\rho+1\\
		&=\frac{4}{p}\left|\{x\in\mathbb{F}_p: x,x+z\in S\}\right|-4\rho+1\\
		&=1-4\rho+\frac{4r_{S-S}(z)}{p}.
	\end{aligned}
	\]
	Let $I=\{0,\ldots,m-1\}$. For this interval, we have $r_{I-I}(0)=m$ and $r_{I-I}(\pm z)=m-z$ for $1\le z<m$, with $r_{I-I}(z)=0$ otherwise. Since $m\leq p/2$, applying Theorem \ref{thm:pollard} to $S$ and $-S$ gives, for every integer $1\leq j\leq m$,
	\begin{align}
		\sum_{z\in\mathbb{F}_p}\min\{j,r_{S-S}(z)\}
		&\geq j\min\{p,2m-j\}
		=j(2m-j)
		=j+2\left(\sum_{k=1}^{j-1}k+\sum_{k=j}^{m-1}j\right)\notag\\
		&=j+2\sum_{k=1}^{m-1}\min\{j,k\}
		=\sum_{z\in\mathbb{F}_p}\min\{j,r_{I-I}(z)\}.
		\label{eq:pollard-interval-comparison}
	\end{align}
	Since $\sum_z r_{S-S}(z)=\sum_z r_{I-I}(z)=m^2$, using the identity $(u-j)_+=\max\{u-j,0\}=u-\min\{u,j\}$ together with \eqref{eq:pollard-interval-comparison}, we obtain $\sum_z(r_{S-S}(z)-j)_+\le\sum_z(r_{I-I}(z)-j)_+$. To extend the preceding inequality to general convex functions, observe that any convex function $\Phi$ on $[0,m]$ satisfies, for integers $0\leq u\leq m$,
	\[\Phi(u)
	=\Phi(0)+u(\Phi(1)-\Phi(0))+\sum_{j=1}^{m-1}
	\big(\Phi(j+1)-2\Phi(j)+\Phi(j-1)\big)(u-j)_+.\]
	The coefficients in the sum are nonnegative by convexity. Applying this identity to $r_{S-S}(z)$ and $r_{I-I}(z)$, and summing over $z$. Since the $r_{S-S}(z)$ and  $r_{I-I}(z)$ have the same total sum, the affine terms contribute equally, while the positive-part terms satisfy the preceding inequalities. Consequently,
	\[\sum_{z\in\mathbb{F}_p}\Phi(r_{S-S}(z))
	\le\sum_{z\in\mathbb{F}_p}\Phi(r_{I-I}(z)) \quad \text{for every convex function} \ \Phi.\]
	Applying this with $\Phi(u)=G(1-4\rho+4u/p)$ gives $\mathbb E_{z\leftarrow\mathbb{F}_p}G(K(z))\le\mathbb E_{z\leftarrow\mathbb{F}_p}G(K_I(z))$, where $K_I(z)=1-4\rho+4r_{I-I}(z)/p$.
		
	To estimate $\mathbb E_{z\leftarrow\mathbb{F}_p}G(K_I(z))$, consider the interval $J=[0,\rho)$ on $\mathbb R/\mathbb Z$, and define $h_\rho=2\mathbbm{1}_J-1$. Let
	\[
	\begin{aligned}
		K_\rho(s)
		&=\int_0^1h_\rho(x)h_\rho(x+s)\,dx=1-4\rho+4\int_0^1\mathbbm{1}_J(x)\mathbbm{1}_J(x+s)\,dx\\
		&=1-4\rho+
		4\big((\rho-s)_++(\rho+s-1)_+\big),
		\qquad 0\le s\le1\\
		&=\begin{cases}
			1-4s, & 0\leq s\leq\rho,\\
			1-4\rho, & \rho\leq s\leq1-\rho,\\
			4s-3, & 1-\rho\le s\leq1.
			\end{cases}
    \end{aligned}
	\]
	Since $r_{I-I}(z)=(m-z)_++(m+z-p)_+$ for $z=0,\ldots,p-1$, we have $K_I(z)=K_\rho(z/p)$. Moreover, $K_\rho$ is $4$-Lipschitz, and so is $G$. Hence $G\circ K_\rho$ is $16$-Lipschitz. Partitioning $[0,1]$ into $p$ equal subintervals, we have $|G(K_\rho(z/p))-G(K_\rho(s))|\le16/p$ whenever $z/p\le s\le(z+1)/p$. Thus,
	\[\left|\mathbb E_{z\leftarrow\mathbb{F}_p}G(K_I(z))
	-\int_0^1G(K_\rho(s))\,ds\right|=\left|\sum_{z=0}^{p-1}
	\int_{z/p}^{(z+1)/p}
	\big[G(K_\rho(z/p))-G(K_\rho(s))\big]\,ds
	\right|\le\frac{16}{p},\]
	Therefore, $\mathbb E_{z\leftarrow\mathbb{F}_p}G(K_I(z))\le\int_0^1G(K_\rho(s))\,ds+16/p$.
	By the piecewise expression of $K_\rho$ and $K_\rho(s)=K_\rho(1-s)$ for $s\in [0,\rho]$, we have
	\[
	\begin{aligned}
		\int_0^1G(K_\rho(s))\,ds
		&=2\int_0^\rho G(1-4s)\,ds
		+(1-2\rho)G(1-4\rho)\\
		&=\frac12\int_{2\beta-1}^1G(u)\,du
		+\beta G(2\beta-1)
		=:T(\beta),
	\end{aligned}
	\]
	where the second equality follows by the substitution $u=1-4s$ and the relation $\beta=1-2\rho$. Consequently, $\mathbb E_{z\leftarrow\mathbb{F}_p}G(K(z))\le T(\beta)+16/p$. Differentiation gives $T'(\beta)=2\beta G'(2\beta-1)$ wherever the derivative exists. Since $G'\le0$ on $[-1,-3/4]$ and $G'=0$ on $[-3/4,3/4]$, the function $T$ is nonincreasing on $[0,1/8]$ and constant on $[1/8,7/8]$. Hence, for $0\le\beta\le7/8$,
	\[T(\beta)\le T(0)=\frac12\int_{-1}^{1}G(z)\,dz
	=\frac{155}{384}.
	\]
		
	Combining the preceding estimates, we conclude that
	\[\|f\|_{U^3}^8\le\frac{155}{384}+\frac{16}{p}+\frac{8}{3p^2}.
	\]
	For $p\ge3000$, the right-hand side is strictly smaller than $41/100$, proving the proposition.  
\end{proof}

\section{Proof of the Statistical Distance Bounds}\label{app:statistical-distance-bounds}

\begin{proof}[Proof of Theorem~\ref{thm:output-decomposition}]

    For convenience, we limit our scope to the case in which all leakage functions $f_i$ are nonconstant. In this case, $|\mu_i|<1$ and $\sigma_i>0$ for every $i$, so all the quantities defined below are well-defined. When some $f_i$ is constant, its output is deterministic and identical under the two distributions being compared. Deleting such coordinates therefore does not change their statistical distance. The following argument can then be applied to the remaining nonconstant coordinates. Since $g_i=h_i=0$ for constant $f_i$, all the corresponding correlation terms vanish, and the desired bound in its original form follows.

    We first prove \eqref{eq:normalized-split}. For each $i\in[n]$, let $ \mu_i:= \mathbb{E}f_i,\ \sigma_i:= \sqrt{1-\mu_i^2}$.  Define the marginal distribution of the $i$-th leakage bit by $q_i(b_i) :=\Pr_{x\leftarrow\mathbb{F}_p}[f_i(x)=b_i]$, where $b_i\in\{-1,1\}$. Since $\mu_i=1\cdot\Pr[f_i(x)=1]+(-1)\cdot\Pr[f_i(x)=-1]$ and $1=\Pr[f_i(x)=1]+\Pr[f_i(x)=-1]$,it follows that $q_i(b_i)=\frac{1+b_i\mu_i}{2}$. For $\mathbf{b}=(b_1,\ldots,b_n)\in\{-1,1\}^n$, define the product distribution $Q(\mathbf{b}):= \prod_{i=1}^n q_i(b_i).$ Following the same normalizing idea used for $g_i$, for each $b_i\in\{-1,1\}$, define
	$\psi_i(b_i):=\frac{b_i-\mu_i}{\sigma_i}$, and for $A\subseteq[n]$, let $\psi_A(\mathbf{b})=\prod_{i\in A}\psi_i(b_i)$, with $\psi_\emptyset=1$. When $b_i$ is distributed according to $q_i$, we have $\mathbb{E}_{b_i\leftarrow q_i}[\psi_i(b_i)]=0$, and $\mathbb{E}_{b_i\leftarrow q_i}[\psi_i(b_i)^2]=1.$ The verification of these identities is given in~\cite[Lemma 12]{rong-github} . Since $Q$ is a product distribution on $\{-1,1\}^n$, its coordinates are mutually independent. Hence, the family $\left\{\psi_A:A\subseteq[n]\right\}$ is orthonormal with respect to $Q$. Specifically, $\mathbb{E}_{\mathbf{b}\leftarrow Q}\left[\psi_A(\mathbf{b})\psi_B(\mathbf{b})\right]
	=\mathbbm{1}_{\{A=B\}}$, 
	a detailed verification is deferred to~\cite[Lemma 13]{rong-github} .
		
	For each \(\mathbf{z}\in\{\mathbf{v},\mathbf{w}\}\), let $P_{\mathbf{z}}(\mathbf{b})=\Pr_{\mathbf{X}\leftarrow\mathbf{z}+C}[\mathbf{f}(\mathbf{X})=b]$, and define \(D_{\mathbf{z}}(\mathbf{b})=P_{\mathbf{z}}(\mathbf{b})/Q(\mathbf{b})\), the density of \(P_{\mathbf{z}}\) with respect to \(Q\). Since \(\psi_i(f_i(x))=g_i(x)\), the coefficient of \(\psi_A\) in the orthogonal expansion of \(D_{\mathbf{z}}\) is
	\[
	\begin{aligned}
		\mathbb E_{\mathbf{b}\leftarrow Q}[D_{\mathbf{z}}\psi_A]
		&=\sum_{\mathbf{b}\in\{-1,1\}^n}Q(\mathbf{b})D_{\mathbf{z}}(\mathbf{b})\psi_A(\mathbf{b})
		=\sum_{\mathbf{b}\in\{-1,1\}^n}P_{\mathbf{z}}(\mathbf{b})\psi_A(\mathbf{b})\\
		&=\mathbb{E}_{\mathbf{b}\leftarrow P_{\mathbf{z}}}\psi_A(\mathbf{b})
		=\mathbb E_{X\leftarrow \mathbf{z}+C}\psi_A(f_1(X_1),\ldots,f_n(X_n))\\
		&=\mathbb{E}_{X\leftarrow \mathbf{z}+C}\prod_{i\in A}\psi_i(f_i(X_i))
		=\mathbb{E}_{X\leftarrow \mathbf{z}+C}\prod_{i\in A}g_i(X_i)
		=\Gamma_{\mathbf{z}}(g_A).
	\end{aligned}
	\]
	By the orthonormality and completeness of \(\{\psi_A\}_{A\subseteq[n]}\), we have \[D_{\mathbf{z}}=\sum_{A\subseteq[n]}\mathbb{E}_{\mathbf{b}\leftarrow Q}[D_{\mathbf{z}}\psi_A]\psi_A=\sum_{A\subseteq[n]}\Gamma_{\mathbf{z}}(g_A)\psi_A.\]
	Consequently, \(D_{\mathbf{v}}-D_{\mathbf{w}}=\sum_{A\subseteq[n]}\left(\Gamma_{\mathbf{v}}(g_A)-\Gamma_{\mathbf{w}}(g_A)\right)\psi_A=\sum_{A\subseteq[n]}a_A^g\psi_A\). By the definition of statistical distance and the preceding orthogonal expansion, we have
	\[
	\begin{aligned}
		&2\mathsf{SD}(\mathbf{f}(\mathsf{share}(s_1)),\mathbf{f}(\mathsf{share}(s_2)))\\
		&=\sum_{\mathbf{b}\in\{-1,1\}^n}|P_{\mathbf{v}}(\mathbf{b})-P_{\mathbf{w}}(\mathbf{b})|
		=\sum_{\mathbf{b}\in\{-1,1\}^n}|Q(\mathbf{b})D_{\mathbf{v}}(\mathbf{b})-Q(\mathbf{b})D_{\mathbf{w}}(\mathbf{b})|\\
		&=\sum_{\mathbf{b}\in\{-1,1\}^n}
		Q(\mathbf{b})|D_{\mathbf{v}}(\mathbf{b})-D_{\mathbf{w}}(\mathbf{b})|
		=\mathbb E_{\mathbf{b}\leftarrow Q}|D_{\mathbf{v}}-D_{\mathbf{w}}|
	    =\mathbb{E}_{\mathbf{b}\leftarrow Q}\left|\sum_{A\subseteq[n]}a_A^g\psi_A\right|\\
		&\le\mathbb E_{\mathbf{b}\leftarrow Q}\left|\sum_{|A|\le r_0}a_A^g\psi_A\right|
		+\sum_{|A|>r_0}|a_A^g|\mathbb{E}_{\mathbf{b}\leftarrow Q}|\psi_A|\\
		&\leq \left(\mathbb{E}_{\mathbf{b}\leftarrow Q}|\sum_{|A|\le r_0}a_A^g\psi_A|^2\right)^{1/2}+\sum_{|A|>r_0}|a_A^g|\mathbb E_{\mathbf{b}\leftarrow Q}|\psi_A|\\
		&=\left(\sum_{|A|\le r_0}|a_A^g|^2\right)^{1/2}
		+\sum_{|A|>r_0}|a_A^g|\mathbb{E}_{\mathbf{b}\leftarrow Q}|\psi_A|.
	\end{aligned}
	\]
	Here, the first inequality follows from the triangle inequality, while the second follows from Cauchy--Schwarz, and the last equality from orthonormality of \(\{\psi_A\}\), which gives \(\mathbb{E}_{\mathbf{b}\leftarrow Q}|\sum_{|A|\le r_0}a_A^g\psi_A|^2
	=\sum_{\substack{|A|\le r_0\\|B|\le r_0}}
	a_A^g\overline{a_B^g}\,
	\mathbb{E}_{\mathbf{b}\leftarrow Q}[\psi_A\psi_B]
	=\sum_{|A|\le r_0}|a_A^g|^2\).
	Since $Q$ is a product distribution and \(\mathbb{E}_{q_i}|\psi_i|
	=\frac{1+\mu_i}{2}\frac{1-\mu_i}{\sigma_i}
	+\frac{1-\mu_i}{2}\frac{1+\mu_i}{\sigma_i}
	=\frac{1-\mu_i^2}{\sigma_i}
	=\sigma_i\), we have \(\mathbb{E}_{\mathbf{b}\leftarrow Q}|\psi_A|=\prod_{i\in A}\sigma_i\). Furthermore, using $h_A=(\prod_{i\in A}\sigma_i)g_A$, we obtain
	\[
	\begin{aligned}
        |a_A^g|\mathbb{E}_{\mathbf{b}\leftarrow Q}|\psi_A|
		&=\left|\Gamma_{\mathbf{v}}(g_A)-\Gamma_{\mathbf{w}}(g_A)\right|
		\prod_{i\in A}\sigma_i
		=\left|
		\Gamma_{\mathbf{v}}\left(\left(\prod_{i\in A}\sigma_i\right)g_A\right)
		-\Gamma_{\mathbf{w}}\left(\left(\prod_{i\in A}\sigma_i\right)g_A\right)
		\right|\\
		&=|\Gamma_{\mathbf{v}}(h_A)-\Gamma_{\mathbf{w}}(h_A)|
		\leq|\Gamma_{\mathbf{v}}(h_A)|+|\Gamma_{\mathbf{w}}(h_A)|
		=\sum_{\mathbf{z}\in{\mathbf{v},\mathbf{w}}}|\Gamma_{\mathbf{z}}(h_A)|.
	\end{aligned}
	\]
	Substituting this bound into the preceding inequality establishes \eqref{eq:normalized-split}.
		
	For \eqref{eq:walsh-split}, consider the uniform distribution $U$ on $\{-1,1\}^n$, and define $\widetilde{D}_{\mathbf{z}}(\mathbf{b})=2^nP_{\mathbf{z}}(\mathbf{b})$ for $\mathbf{z}\in\{\mathbf{v},\mathbf{w}\}$. The Walsh functions $b_A=\prod_{i\in A}b_i$, $A\subseteq[n]$, form an orthonormal basis of $U$. By the same argument as above, the expansion coefficients of $\widetilde D_{\mathbf{z}}$ are $\mathbb{E}_{\mathbf{b}\leftarrow U}[\widetilde D_{\mathbf{z}}b_A]=\Gamma_{\mathbf{z}}(f_A)$. Consequently,
	\[
	\begin{aligned}
		2\mathsf{SD}(\mathbf{f}(\mathsf{share}(s_1)),\mathbf{f}(\mathsf{share}(s_2)))
		&=\mathbb{E}_{\mathbf{b}\leftarrow U}|\widetilde{D}_{\mathbf{v}}-\widetilde{D}_{\mathbf{w}}|
		=\mathbb{E}_{\mathbf{b}\leftarrow U}\left|\sum_{A\subseteq[n]}a_A^fb_A\right|\\
		&\leq\left(\mathbb{E}_{\mathbf{b}\leftarrow U}\left|\sum_{A\subseteq[n]}a_A^fb_A\right|^2\right)^{1/2}
		=\left(\sum_{A\subseteq[n]}|a_A^f|^2\right)^{1/2}.
	\end{aligned}
	\]
	This establishes~\eqref{eq:walsh-split} and completes the proof.
\end{proof}
    
\section{Proof of the Operator Norm Estimate}\label{app:normalized-low-support-transfer}
	
We establish the operator norm estimate used in Lemma \ref{lem:normalized-low-support-transfer}.
	
\begin{lemma}
\label{lem:operator-norm}
	Let $T=u^2\in(1,2)$ and $x=\mu^2\in[0,1]$. Define the squared one-coordinate operator norm by
	\[\lambda_T(x):=
	\left\|
	\begin{pmatrix}
		1&0\\
		\mu&\sqrt{1-\mu^2}
	\end{pmatrix}
	\text{diag}(1,u)
	\right\|_{\text{op}}^2.
	\]
	Then, we have
	\begin{equation}
	\label{eq:s7-lambda}
		\lambda_T(x)=\frac{1+T+(1-T)x+\sqrt{(T-1)^2(1-x)^2+4x}}{2},
	\end{equation}
	and as a function of $x$, $\lambda_T$ is positive, concave, and nondecreasing on $[0,1]$. Consequently, if $B\le B_0\le1$, then
	\begin{equation}
	\label{eq:s7-product-norm}
		\prod_{i=1}^n \|M_i\text{diag}(1,u)\|_{\text{op}} \le\lambda_T(B)^{n/2} \le\lambda_T(B_0)^{n/2}.
	\end{equation}
\end{lemma}
\begin{proof}
	For a single coordinate, write
	\[A=
    \begin{pmatrix}
		1&0\\
		\mu&\sigma
	\end{pmatrix}\mathrm{diag}(1,u)
	=
	\begin{pmatrix}
		1&0\\
		\mu&u\sigma
	\end{pmatrix},
	\qquad \sigma=\sqrt{1-\mu^2}.
	\]
	Since the squared operator norm equals the largest eigenvalue of $A^{\mathsf T}A$, where
	\[
	A^{\mathsf T}A
	=
	\begin{pmatrix}
		1+\mu^2&u\mu\sigma\\
		u\mu\sigma&u^2\sigma^2
	\end{pmatrix}.
	\]
	The eigenvalues of $A^{\mathsf T}A$ are the roots of
	\[
	\lambda^2-\bigl(1+T+(1-T)x\bigr)\lambda+T(1-x)=0,
	\]
	Then, the larger root is precisely $\lambda_T(x)$ given in \eqref{eq:s7-lambda}. Moreover,
	\[
	\lambda_T(x)\geq \frac{1+T+(1-T)x}{2}=\frac{2+(T-1)(1-x)}{2}\geq1>0
	\]
	by $T>1$ and $x\in [0,1]$.
		
	Next, we show that $\lambda_T(x)$ is concave and nondecreasing on $[0,1]$. We calculate the deritive and the second derivative of $\lambda_{T}(x)$ directly:
	\begin{align*}
		\lambda^{'}_{T}(x)& =\frac{1-T}{2}+\frac{2-(T-1)^2(1-x)}{2\sqrt{(T-1)^2(1-x)^2+4x}},\\
		\lambda^{''}_{T}(x)& =\frac{2T(T-2)}{((T-1)^2(1-x)^2+4x)^{3/2}}.
	\end{align*}
	Since $T\in (1,2)$, we have $\lambda^{''}_{T}(x)<0$ for $x\in [0,1]$. Note that $\lambda_T'(1)=\frac{2-T}{2}>0$. Hence, concavity makes $\lambda_T'$ nonincreasing. Thus, for every $x\in[0,1]$, we have $\lambda_T'(x)\ge\lambda_T'(1)>0$, which shows that $\lambda_T(x)$ is nondecreasing.
		
	Since $\lambda_T$ is positive and concave, $\ln\lambda_T$ is also concave. Writing $x_i=\mu_i^2$, by Jensen's inequality, we have
	\[\frac{1}{n}\sum_{i=1}^n\ln\lambda_T(x_i)\le\ln\lambda_T\left(\frac{1}{n}\sum_{i=1}^n x_i\right)=\ln\lambda_T(B).\]
	Consequently,
	\[\prod_{i=1}^n	\|M_i\text{diag}(1,u)\|_{\text{op}} =\exp\left(\frac12\sum_{i=1}^n\ln\lambda_T(x_i)\right) \le\lambda_T(B)^{n/2} \le\lambda_T(B_0)^{n/2},\]
	where the last inequality uses $B\le B_0$ and the monotonicity of $\lambda_T(x)$.
\end{proof}
	
\section{Proof of the Boolean Product Decorrelation Estimate}\label{app:boolean-product-decorrelation}

We establish the Boolean product decorrelation bound used in Proposition \ref{prop:large-support-small-bias}. Our proof uses the Fourier thresholding method of Sun and Wootters \cite[Lemma 1.15]{SunWootters2026}
	
\begin{proof}[Proof of Lemma~\ref{lem:decor}]
\label{lem:decor proof}
	If $d=m$, then $C=\mathbb{F}_p^m$ and its coordinates are fully independent, so the left-hand side is exactly zero.  We may therefore assume $d<m$ for the rest of the proof. 
		
	Expanding each $f_i$ in the Fourier basis and applying character orthogonality, we obtain
	\[ \mathbb E_{\mathbf{X}\leftarrow C}\prod_{i=1}^m f_i(X_i) =\sum_{\mathbf{y}\in C^{\perp}} \prod_{i=1}^m\widehat{f_i}(y_i).\]
	The all-zero word contributes $\prod_i\widehat{f_i}(0)=\prod_i\mathbb{E}f_i$.  After subtracting it, we obtain
	\begin{equation}
		\left|\mathbb E_{\mathbf{X}\leftarrow C}\prod_i f_i(X_i)
		-\prod_i\mathbb E f_i\right|
		\le\sum_{\mathbf{0}\ne\mathbf{y}\in C^{\perp}}
		\prod_{i\in\mathsf{supp}(\mathbf{y})}
		|\widehat f_i(y_i)|.
		\label{eq:appc-fourier-sum}
	\end{equation}
	According to Lemma \ref{lem:leakage-fourier}, and Parseval's identity, for each $i\in[m]$, we have  $ |\widehat {f_i}(a)|\le\gamma, a\ne0$, and
	$\sum_{a\in\mathbb{F}_p}|\widehat {f_i}(a)|^2=\mathbb{E}|f_i|^2=1.$
		
	Fix $\theta=(\gamma/3)^{10}$ and define the lists $H_i=\{0\}\cup\{a\neq0: |\widehat {f_i}(a)|>\theta\}.$ Each list has size at most $1+\theta^{-2}$. We call the $i$-th coordinate of a codeword in $C^{\perp}$ light, if it does not belong to the list $H_i$. We split the sum in~\eqref{eq:appc-fourier-sum} according to the number of light coordinates.
		
	First consider nonzero words of $C^{\perp}$ with at most $\lfloor m/10\rfloor$ light coordinates. These codewords agree with the lists on at least $m-\lfloor m/10\rfloor$ coordinates. Since $C^{\perp}$ is a GRS of dimension $m-d\le m/11$, for sufficiently large $m$,
	\[ (m-d)+\left\lceil\frac{4}{5}m\right\rceil \le m-\left\lfloor\frac {m}{10}\right\rfloor.\]
	By Theorem \ref{thm:rs-list-recovery}, with $\xi=4/5$ and constant list size, the number of such codewords is at most $m^{O(1)}$. Since every nonzero word of $C^{\perp}$ has weight at least $d+1$, so their total contribution is at most $m^{O(1)}\gamma^{d+1}$.
		
	Every remaining codeword has more than $0.1m$ nonzero light coordinates. Let $T$ be its support and choose $M\subseteq T$ of size $\lceil0.1m\rceil$. By Lemma \ref{lem:grs-mds}, the shortened dual on $T$ has dimension $|T|-d$, and every $|T|-d$ coordinates are an information set. Since $|T|\le m$ and $m\leq 1.1d$, for sufficiently large $m$,
	\[|M|\le0.1m+1\le2d-|T|,\]
	which is equivalent to 
	\[2(|T|-d)\le |T|-|M|.\]
	So $T\setminus M$ contains two disjoint subsets $I_1,I_2$ of size $|T|-d$, and each $I_j$
	is therefore an information set.

    Fix the pair $(T,M)$ and denote $D_T=(C^\perp)_T$, so that $\dim D_T=|T|-d$. Let
    \[
    \mathcal E_{T,M} = \left\{\mathbf y\in C^\perp:\mathsf{supp}(\mathbf y)=T,\ y_i\notin H_i\ \text{for all }i\in M \right\}.
    \]
    For every $\mathbf y\in\mathcal E_{T,M}$, the coordinates indexed by $M$ are light, and hence $|\widehat{f_i}(y_i)|\le\theta$ for all $i\in M$. Moreover, all Fourier coefficients have magnitude at most $1$. Therefore,
    \begin{align*}
    \sum_{\mathbf y\in\mathcal E_{T,M}}\prod_{i\in T}|\widehat{f_i}(y_i)| \le 
    \theta^{|M|} \sum_{\mathbf y\in\mathcal E_{T,M}} \prod_{i\in I_1\cup I_2} |\widehat{f_i}(y_i)|
    \le \theta^{|M|} \sum_{\mathbf y\in D_T} \prod_{i\in I_1\cup I_2} |\widehat{f_i}(y_i)|,
    \end{align*}
    where the second inequality follows by enlarging the summation to all codewords in $D_T$, identifying each shortened codeword with its extension by zeros outside $T$.
    
    By the Cauchy--Schwarz inequality, we have
    \begin{align*}
    \sum_{\mathbf y\in D_T} \prod_{i\in I_1\cup I_2} |\widehat{f_i}(y_i)|
    & \le \left(\sum_{\mathbf y\in D_T}\prod_{i\in I_1}|\widehat{f_i}(y_i)|^2\right)^{1/2}
    \cdot
    \left(\sum_{\mathbf y\in D_T}\prod_{i\in I_2}|\widehat{f_i}(y_i)|^2\right)^{1/2}=1.
    \end{align*}
    Here, the last equality follows because the projection onto either information set $I_j$, $j=1,2$, is a bijection, and Parseval's identity gives
    \begin{align*}
    \sum_{\mathbf y\in D_T}\prod_{i\in I_j}|\widehat{f_i}(y_i)|^2 = \prod_{i\in I_j} \sum_{a\in\mathbb F_p}|\widehat{f_i}(a)|^2=1.
    \end{align*}
    Consequently,
    \[
    \sum_{\mathbf y\in\mathcal E_{T,M}}\prod_{i\in T}|\widehat{f_i}(y_i)| \le \theta^{|M|}.
    \]
    
    Every codeword in the second class belongs to at least one such set $\mathcal E_{T,M}$. There are at most $3^m$ admissible pairs $(T,M)$, since each coordinate belongs to one of $M$, $T\setminus M$, and $[m]\setminus T$. Since $|M|=\lceil m/10\rceil$, the total contribution from the second class is bounded by
    \[
    3^m\theta^{\lceil m/10\rceil}
    \le 3^m\theta^{m/10}=\gamma^m \le \gamma^d.
    \]
    Combining the two classes, we obtain
    \begin{align*}
    \left|
    \mathbb E_{\mathbf X\leftarrow C} \prod_{i=1}^m f_i(X_i) -\prod_{i=1}^m\mathbb E f_i 
    \right| \le
    \sum_{\mathbf 0\ne\mathbf y\in C^\perp}\prod_{i:y_i\ne0}|\widehat{f_i}(y_i)|
    \le m^{O(1)}\gamma^{d+1}+\gamma^d \le m^{O(1)}\gamma^d.
    \end{align*}
    
    The above arguments apply for all sufficiently large $m$. For the finitely many smaller values of $m$, the left-hand side is at most $2$. Since $d\le m$, the desired estimate continues to hold after enlarging the absolute implicit constant, independently of $p$.
    
    For affine translates, the shift can be absorbed into the Boolean functions without changing their means. Thus, the same bound holds. This completes the proof.
        
\end{proof}

\end{document}